\documentclass{aa}  
\usepackage{graphicx}
\usepackage{subfigure}
\usepackage{txfonts}
\def\arcm{\hbox{$^\prime$}}
\def\arcs{\arcm\hskip -0.1em\arcm}
\begin{document}
   \title{The XMM-Newton EPIC Background and the production of Background Blank Sky Event Files \thanks{Based on observations with XMM-{\it Newton},
an ESA Science Mission with instruments and contributions directly
funded by ESA Member States and the USA (NASA).}}

   \subtitle{}

   \author{J.A. Carter
          \inst{1}
          \and
          A.M. Read\inst{1}
          }

   \offprints{J.A. Carter}

   \institute{Department of Physics and Astronomy, University of Leicester, 
 Leicester, LE1 1RH, UK\\
              \email{jac48@star.le.ac.uk}\\
	      \email{amr30@star.le.ac.uk}\\
                }

   \date{Received 22 June 2006; accepted 22 December 2006}

 
  \abstract
   {}
   {We describe in detail the nature of XMM-Newton EPIC background and its various complex components, summarising the new findings of the XMM-Newton EPIC background working group, and provide XMM-Newton background blank sky event files for use in the data analysis of diffuse and extended sources.}
   {Blank sky event file data sets are produced from the stacking of data, taken from 189 observations resulting from the Second XMM-Newton 
Serendipitous Source Catalogue (2XMMp) reprocessing. The data underwent several filtering steps, using a revised and improved method over previous work, which we describe in detail.}
   {We investigate several properties of the final blank sky data sets. The user is directed to the location of the final data sets. There is a final data set for each EPIC instrument-filter-mode combination.}
   {}


   \titlerunning{XMM-Newton EPIC blank sky event files}
   \authorrunning{J.A. Carter \& A.M. Read}
   \maketitle
%

\section{Introduction}

The XMM-Newton observatory (Jansen et al. (2001)), with the two EPIC
MOS (Turner et al. (2001)) and one EPIC PN (Str\"{u}der et al. (2001))
cameras at the foci of the three telescopes, provides unrivalled
capabilities for detecting low surface brightness emission features
from extended and diffuse galactic and extragalactic sources, by
virtue of the large field of view of the X-ray telescopes and the high
throughput yielded by the heavily nested telescope mirrors. In order
to exploit the excellent EPIC data from extended objects, the EPIC
background (BG), now known to be higher than estimated
pre-launch, needs to be understood thoroughly.

In 2005 a working group was set up to organise
various investigations into the BG, to define the BG material useful
for the general user and to place this BG material and findings of the
group on the main XMM-Newton SOC web pages. This paper concentrates on
one aspect of this work, namely that of the production of blank sky
background event files, constructed for each of the EPIC instruments
in each filter/mode combination.

The work described in this paper follows on from that of Read \&
Ponman (2003). The number of observations used in this analysis has
increased from 72 to 189, greatly improving the signal-to-noise
ratio. Further changes to the previous dataset have been the inclusion
of thick filter observations in the analysis and other improvements
described later.

All the background product files (event files, exposure maps, related
software), together with other scripts and procedures for XMM-Newton
background analysis are available from the official ESA site: \newline
\indent \textit{http://xmm.vilspa.esa.es/external/}\newline \indent
\indent \textit{xmm\_sw\_cal/background/index.shtml.}

The activities of the BG working group has led to a greater
understanding of the complex components that make up the XMM-Newton
EPIC BG. In section 2 of this paper we describe these various
components, presenting a review of our current understanding of the
BG. In section 3 we describe the production of the blank sky event
files, detailing improvements made over previous work in the filtering
stages using current knowledge of the BG. In section 4 we detail some
properties of these files, and in section 5 we discuss the use of
these files and present software that we have written and made
available for use with the blank sky event files. Finally, in section
6 we present our conclusions and discuss future plans for analysing
the BG.

\section{The XMM-Newton EPIC X-ray Background} 

Given the complex nature of the EPIC background, and that new
components of the background have recently been discovered by the BG
working group and others, it is appropriate that a comprehensive
description of the EPIC background should be given:

The EPIC background can be separated into particle, photon and
electronic noise components (as described in the work of Lumb et
al. (2002) and Read \& Ponman (2003) and references therein). Several
contributions are focused by the mirrors, whereas others arrive at the
detectors directly through the shielding. The particle background can
be further sub-divided into contributions from soft protons and
cosmic-ray induced events, and the photon background can be
sub-divided into contributions from hard and soft X-rays.  In this
section we describe each of these components, discussing their
temporal, spectral and spatial properties. A table summarising the
components of the background can be found via the official XMM-Newton
pages.

\subsection{Particle background}

The particle background consists of focused soft protons and unfocused
cosmic ray induced events.

\subsubsection{Soft protons}

This contribution to the background comes from solar soft protons,
accelerated by magneto-spheric reconnection events and trapped by the
Earth's magnetosphere, which are then gathered by XMM-Newton's grazing
mirrors. They dominate times of high background.  These soft protons
occur in flares up to 1000\% of the quiescent level in an observation.
They are highly unpredictable and affect 30\% to 40\% of XMM-Newton
observation time. The frequency of soft protons seen increases
closer to perigee.  Within a single observation, a significant
component may survive after good time interval screening (de Luca \&
Molendi 2004). Spectrally the soft protons are variable in
intensity and shape. For energies $>$ 0.5\,keV the continuum spectrum,
which shows no lines, can be fitted by an unfolded Xspec power law,
i.e. one not convolved with the instrumental response (specifically a
double-exponential or broken power law, with the break energy at
approximately 3.2\,keV, and with the spectrum becoming flatter at
higher intensities). Below 0.5\,keV, much less flux is seen (Kuntz et
al. in preparation). The soft protons are distributed over the
detector in a similar manner to X-rays, but the vignetting function is
flatter than for photons. Also for the low-energy soft protons the
vignetting function is flatter than that of high-energy protons. They
are only observed inside the FOV (though during times of intense solar
flares (which are unfocused), the out-FOV signal can be greatly
increased). There is no other spatial structure seen in the PN, but
some structure may occur in the MOS due to the presence of the
Reflection Grating Array on board XMM-Newton. An as yet poorly
understood soft proton feature is seen in MOS CCD-2 at low energies
(Kuntz et al. in prep.).

\subsubsection{Internal cosmic ray induced events: the instrumental BG}\label{bginstr}

This component of the BG results from high-energy particles producing
charge directly in the CCDs, and from the interaction of high energy
particles with the detector, causing associated instrumental
fluorescence.  Within an observation this component can vary by up to
10\%. For the MOS cameras above 2\,keV there is no change seen in the
continuum and only small changes seen in the fluorescence lines, but
below 1.5\,keV the continuum varies, possibly due to the
redistribution of the Al calibration line (de Luca \& Molendi
2004). From observation to observation there is some variation; up to
10 times more intense an effect can be seen during periods of intense
solar flares, but no increase is seen after the occurrence of solar
flares so activation is unlikely.

The continuum spectrum is flat (with an index $\sim$0.2). The
instrumental fluorescence lines for MOS are found at 1.5\,keV (Al-K),
1.7\,keV (Si-K), plus some contribution from high energy lines
(Cr, Mn, Fe-K and Au). For the PN, Al-K is seen at 1.5\,keV, whereas
the silicon line is self-absorbed, and high energy contributions are
seen from Cu, Ni, Zn and K. Detector noise occurs below 0.3\,keV.

The internal instrumental BG has a spatial distribution different from
that of X-ray photons as it is not vignetted.  In the outer CCDs and
outside the field of view (FOV) for MOS there is more Al seen than Si,
whereas the CCD edges show enhanced Si. There are continuum
differences between the out-FOV and in-FOV below the Al-line,
possibly resulting from redistribution. There is more Au seen
out-FOV due to the Al-shielding which is coated with gold on its
inner surfaces. Energies and widths of the lines are stable, whereas
line intensities can vary.  In the PN, line intensities show large
spatial variation from the electronics board, for example the
well-known 'copper hole', where a deficit in high-energy instrumental
lines is seen at the detector centre (Freyberg et
al. (2004)). Residual low-energy instrumental BG components are seen
near the CAMEX readout areas.

\subsection{Electronic noise}

Electronic noise results from bright pixels and parts of columns,
CAMEX readout noise in the PN and artificial low-energy enhancements
in the outer CCDs of MOS. Dark current may also contribute, but this
is thought negligible.

No temporal variations are seen within an observation apart from the
bright pixel and column component that can vary by up to 10\%. Bright
pixels fluctuate greatly between observations. For the PN, the CAMEX
readout noise is mode dependent; extended-full-frame mode suffering
the least from this noise, and small window mode the most. Artificial
low-energy enhancements may affect up to 20\% or more of observations,
and are enhanced during periods of high BG rate.

Spectrally, this component is seen at low energies (below 300\,eV) for
the bright pixel and CAMEX readout contributions. Artificial
low-energy enhancements are sometimes observed below 500\,eV in
certain MOS CCDs (Kuntz et al. in prep., Pradas \& Kerp (2005)).

This component is distributed differently from genuine X-ray photons
and is not vignetted, apart from the artificial low-energy
enhancements in the MOS cameras. The bright pixels and columns are
seen at certain locations; structure is seen near to the PN readout
(CAMEX). Certain MOS CCDs show some peculiarities in and out of FOV
(Kuntz in prep.) and spatial inhomogeneities are seen within a
particular MOS CCD.

\subsection{Photon background}

The photon background can be split into components from hard and soft
X-rays and these components are focused by the mirrors.

\subsubsection{Hard X-ray photons}

The hard X-ray background photons mainly originate from unresolved AGN
within the FOV. There are also single reflections into the FOV from
all kinds of out-FOV sources, both bright and faint, resolved and
unresolved (the unresolved out-FOV sources being, as for the in-FOV,
predominatly AGN). Out-of-time events (OOT) are also a contributor to
the hard X-ray BG of the PN.

The hard X-ray photon BG does not vary within an observation or
between observations, although OOT events are mode dependent for the
PN; the full-frame mode experiencing more of this effect than both the
extended-full-frame and large window mode, due to the percentage of
the frame time used for readout.

The hard X-ray photon BG can be modelled by a power law of spectral
index $\sim$1.4. In times of low-BG, and below 5\,keV, this component
dominates over the internal component of the BG, whereas above 5\,keV,
the internal component dominates. As they are genuine X-ray photons,
they are spatially vignetted.

Diffuse flux from single reflections gathered from out-of-field angles
of 0.4$-$1.4 degrees that are reflected into the FOV ('single
reflections'), contribute $\sim$7\% of the in-FOV flux (Lumb et
al. (2002)), and the effective area of one of the telescopes is
approximately 3 cm$^2$ at 20-80\arcm \,off-axis (Freyberg et
al. (2004)). OOT events are smeared along the readout direction from
the bright X-ray sources of X-rays (Freyberg et al. (2004)).

\subsubsection{Soft X-ray photons}

Soft X-rays originate from the Local Bubble, Galactic Disk, Galactic
Halo, the Solar Wind Charge Exchange (SWCX) (Snowden et al. 2004)
single reflections from outside the FOV and OOT events (PN only).  The
SWCX is an interaction between the highly ionised solar wind and
either interstellar neutrals in the heliosphere or materials in the
Earth's exosphere. There is little variation seen in the soft X-ray BG
during a single observation, although long observations may be
affected by the SWCX. Variations of up to 35\% are seen between
observations as observation pointings differ in Right Ascension and
Declination. Also, the SWCX component may effect observations
differently.

The diffuse contributions from the Local Bubble, Galactic Disk and
Galactic Halo have a thermal component with emission lines
$\sim$$<$1\,keV.  The extragalactic component above 0.8\,keV has an
index of 1.4, whereas the galactic contribution in terms of emission
and absorption varies. The SWCX component is very soft and
comprises unusual OVIII/OVII line ratios and strong OVIII and
MgXI features.

The soft X-ray BG component is vignetted as it is made up of genuine
X-ray photons. Spatially, the only structure seen is from real
astronomical objects, and the extragalactic component above 0.8\,keV is
spatially uniform. The SWCX is seen over the whole FOV. The single
reflections and OOT events behave as those resulting from hard
X-rays. \\

The BG has shown itself to be extremely complicated and made up of
various different components. When performing detailed XMM-Newton EPIC
analysis, a good knowledge of the background is required. Sometimes it
may be possible to extract the background from a region close to the
particular source one is interested in (using a so-called `local'
background). For a large or extended source however, one may have to
extract the background far from the target source (the source may in
fact be so extended, that no local background is visible within the
field of view). Here, a number of effects, due to many of the features
described above, can cause the extracted local (off-axis) background
to be highly inappropriate in analysing the (normally on-axis located)
target source, such as changes in the effective area of the mirrors
with off-axis angle, instrumental fluorescence and the spectral
response which can depend on the position on the detector (these
off-axis effects are corrected in the XMM-Newton EPIC
calibration). Hence the need for blank sky background event files for
the general user to study diffuse and extended sources, where images
from XMM-Newton do not provide suitable selection areas for background
subtraction.

\section{Blank sky analysis} 

The data processing described here has resulted in the production of
new XMM-Newton background event files for the 3 EPIC instruments in
their different instrument mode/filter combinations. These have been
constructed using a superposition of many pointed observations of
pipeline product data from the Second XMM-Newton Serendipitous Source
Catalogue (2XMMp) reprocessing: \newline
(\textit{http://xmm.esa.int/external/xmm\_user\_support/} \newline
\indent \textit{documentation/uhb/node141.html}) \newline and have
been processed with the latest version of the SAS, SAS 6.5.0.

These files are available at the website location as mentioned in 
the introduction. 

In the case of PN, in full-frame mode, for the medium and thin filter,
it was necessary to split the event files, as the size of these files
exceeded that of the maximum that is usable with FTOOLS and the
XMM-Newton Science Analysis System (SAS). This is explained in more
detail below.

In total 189 observations were analysed from the 2XMMp
reprocessing. The observations were selected based on the absence of
a large diffuse component or significantly bright source whose wings
could still contaminate the background after source
subtraction. Observations with bright central sources were avoided.
Therefore observations of fields such as the Lockman Hole, Hubble Deep
Field North and Marano pointings etc. were of particular use, and the
archive was scanned for useful survey fields. All available
observations that were used by previous background studies (Read and
Ponman (2003), Nevalainen et al. (2005), Lumb et al. (2002)) were
included.

Table \ref{allobstable} lists the observations used in this analysis. 

The final event lists result from the stacking of pipeline product
event lists from many observations that have been subjected to various
filtering steps, which includes the removal of sources (as described
below). Therefore proper consideration of the exposure maps is
required when using the final event list that applies to a set of
combined observations.

Each observation was subjected to the same initial analysis. The steps
in this procedure were as follows:

\begin{itemize}

\item The relevant 2XMMp pipeline processing system products (PPS)
(event lists, attitude files, background time series, source files and
calibration index files) are collected together.
 
\item For each instrument, region files are created from the PPS
source lists. These are then used to remove all the source events from
each of the relevant event files. A conservative extraction radius of
35\arcs\ is used to remove the sources (for comparison, Read and
Ponman (2003) used 36\arcs\ and Lumb et al. (2002) used 25\arcs).
These regions are also removed from previously created mask files
(these are required to calculate losses in area due to source
removal).

\item A visual inspection is made of the data to make sure that there
are no strange features in the field (such as the rings seen from
off-axis single reflections), and to ascertain whether there are any
wings of very bright point sources or large diffuse sources which
could contaminate the background, even after source
subtraction. Datasets which fail this inspection are rejected from any
subsequent analysis.

\item The event files are then filtered further. Events with energies
below 100\,eV are discarded. For PN, only singles and doubles are
retained, for MOS 1 and MOS 2, singles, doubles, triples and
quadruples are retained. Finally, the event lists are filtered using
the SAS-recommended \textit{\#XMMEA\_EM} and \textit{\#XMMEA\_EP} FLAG
expressions, excepting that events from outside the field of view
(out-FOV) are also kept.

\item Each of the event files are then conservatively filtered for
periods of high background (solar proton flares) by first creating
Good Time Interval (GTI) files from the pipeline processing products
background time series files and then applying these GTI files to the
event file. Upper count rate threshold of 60 (PN) and 2 (MOS1/MOS2)
ct/s are used as recommended as conservative thresholds by the EPIC
instrument teams.

\item In the case of PN, it was necessary to further filter the files
to clean a small number of persistent bad (bright) pixels/columns,
occurring in many (though not all) of the observations. Events were
removed from all PN datasets below 250\,eV, as follows: CCD1 col.13 \&
pixels RAWX,Y=(56,75), CCD2 RAWX,Y=(46$-$47,69$-$72), CCD5 col.11 \&
RAWX,Y=(41, 182-184), CCD7 col.34, CCD10 col.61, CCD11
RAWX,Y=(47$-$48,153$-$156) \& RAWX,Y=(50,161$-$164). No bad pixels
were removed from any of the MOS datasets.  The event files are now
filtered and have had all sources removed.

\item Ghosting of events, if applicable, is applied by an IDL code
(see section \ref{ghosting}). The use of this procedure results in
\textit{refilled} event list, after the completion of the tasks below.
Therefore there are two types of final files: \textit{refilled} files
that have gone through the ghosting procedure, and \textit{unfilled}
files that have not.

\item For each of the three instruments, a non-vignetted and vignetted
4\arcs exposure map is created.  For the unfilled sets, the procedure
was more complex than that of the filled sets, as the sources removed
had to be taken into account. From the source-removed mask file, an
area map (4\arcs binning) is created, containing zero values at the
positions where sources have been removed, and unity values elsewhere.
This is combined with the 4\arcs exposure map to create a source
removed exposure map.

\item To each event, values of Right Ascension and Declination are
given in newly formed columns, using the information from the original
event file header \textit{RA\_PNT} and \textit{DEC\_PNT} keywords.

\end{itemize}

\subsection{Ghosting of events}\label{ghosting} 

\begin{figure}[h]
     \centering
          \includegraphics[width=.23\textwidth]{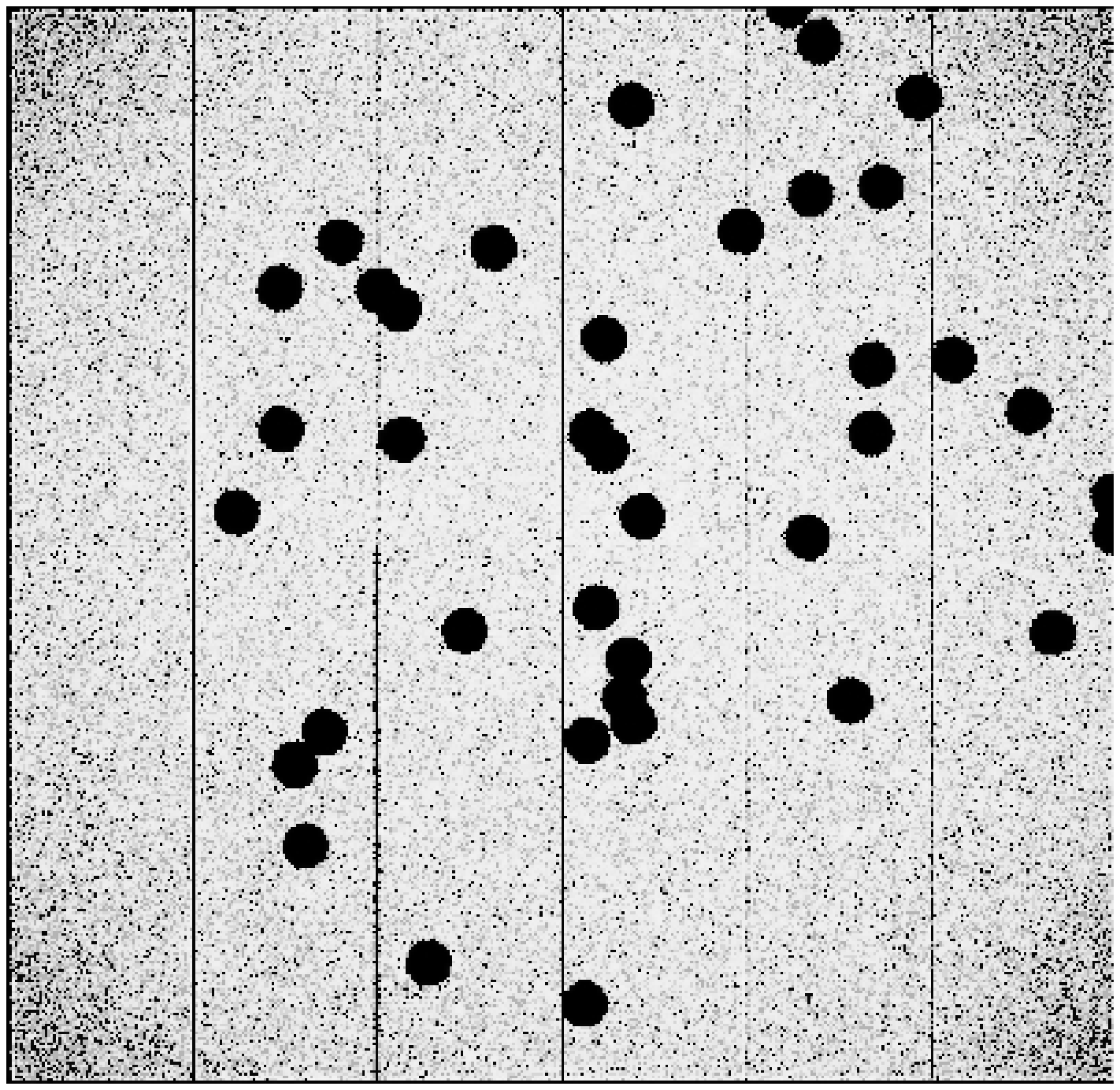}
          \includegraphics[width=.23\textwidth]{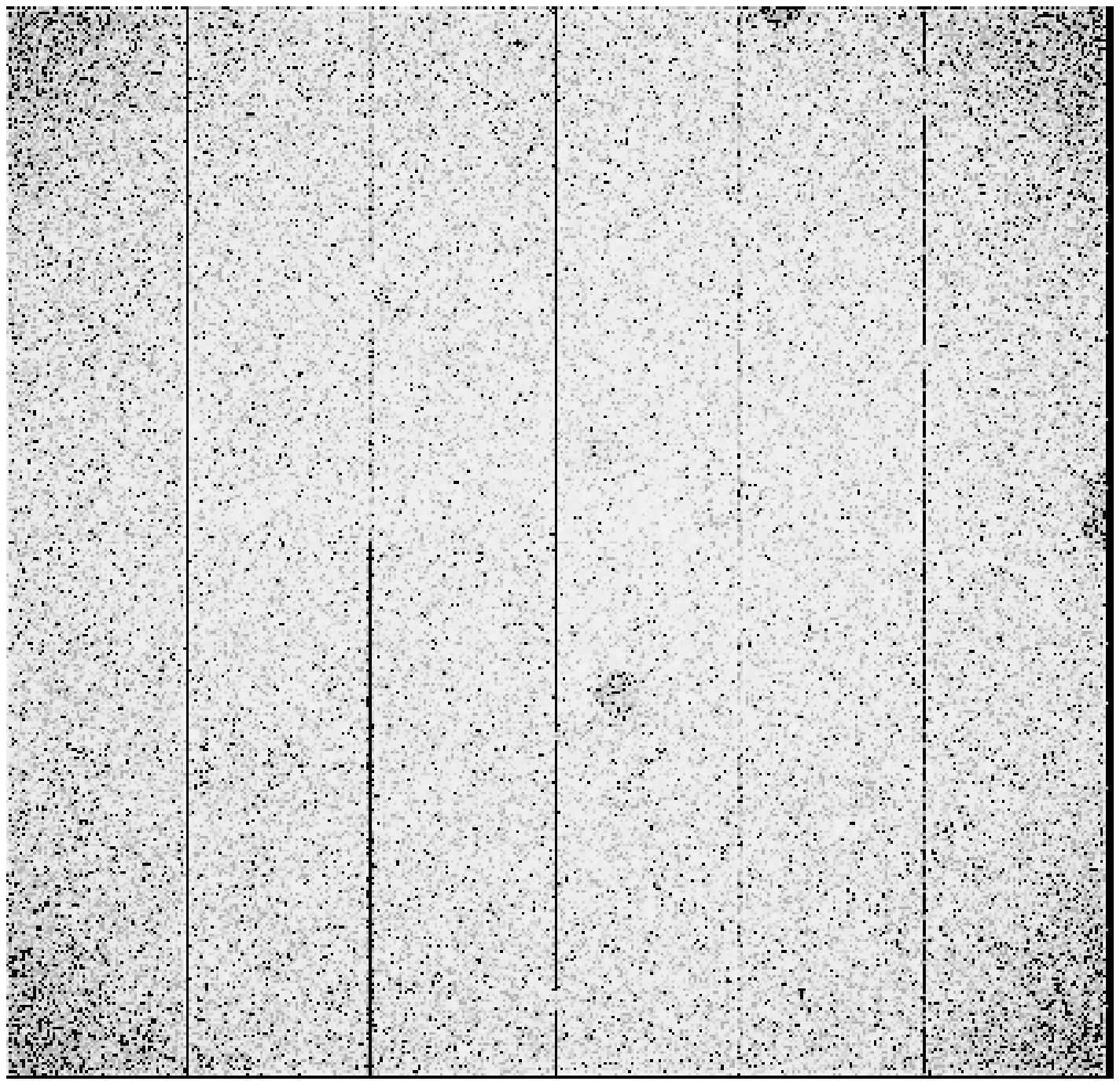}
      \caption{Images from a PN observation with the source holes
        removed prior to the ghosting procedure (left) and after the
        ghosting procedure (right). Although there are slight problems
        in the ghosting procedure in very source confused areas, no
        problems remain after stacking of several such filled event
        files.}
      \label{figghost}
\end{figure}

\begin{figure}
\vspace{1cm}
\includegraphics[height=8cm, width=8cm, angle=180]{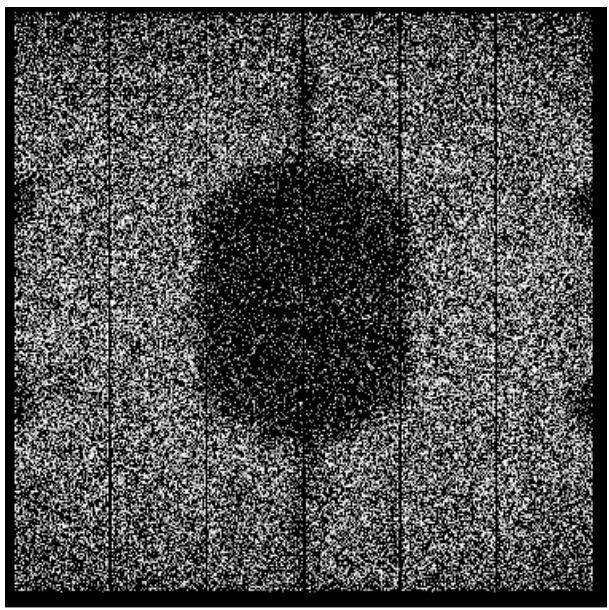}
\caption{An image created from one of the final refilled event files
for PN, thin filter, full-frame mode, between the energies of 7.8\,keV
and 8.2\,keV. This shows the copper hole and the slight blurring of
the edges of this feature due to the ghosting procedure.}
\label{figpn8keV}
\end{figure}

There are two types of background event lists; unfilled and refilled.
In the case of the refilled event lists, a method has been developed
to 'fill in' the source regions that are extracted from each
individual observation by sampling events close to the extracted
regions, copying these events and filling the vacated region of the
event list, randomising just the spatial(\textit{DETX, DETY})
coordinates. Great care was taken when making adjustments for region
crossovers, instrument boundaries and chips edges. This results in
smooth event file images and exposure maps.  Both types of event file
are available at the aforementioned website, with corresponding
vignetted and non-vignetted exposure maps.  As the ghosted events in
this code are based on \textit{DETX, DETY} coordinates, it is
necessary to use the SAS task \textit{attcalc}, setting right
ascension, declination and pointing angle to (0, 0, 0) to re-project
these events in terms of \textit{X} and \textit{Y}.  Figure
\ref{figghost} shows an example individual observation prior to and
post the ghosting procedure. Figure \ref{figpn8keV} shows an image
created from one of the final PN files made up of event files that
have been subjected to the ghosting procedure, between 7.8\,keV and
8.2\,keV, which illustrates that the ghosting procedure slightly blurs
the edges of the 'copper hole' as described in section \ref{bginstr},
but is still of minimal significance. It is hoped in the future that
we will be able to develop the ghosting procedure to ghost bad columns
and pixels (see Section \ref{concl}).

\subsection{Properties of the observations used}\label{propobs} 

\begin{figure*}[htp]
     \centering
     \subfigure[MOS1 livetime histogram after cleaning, all observations]{
          \label{liveclean}
          \includegraphics[width=.45\textwidth]{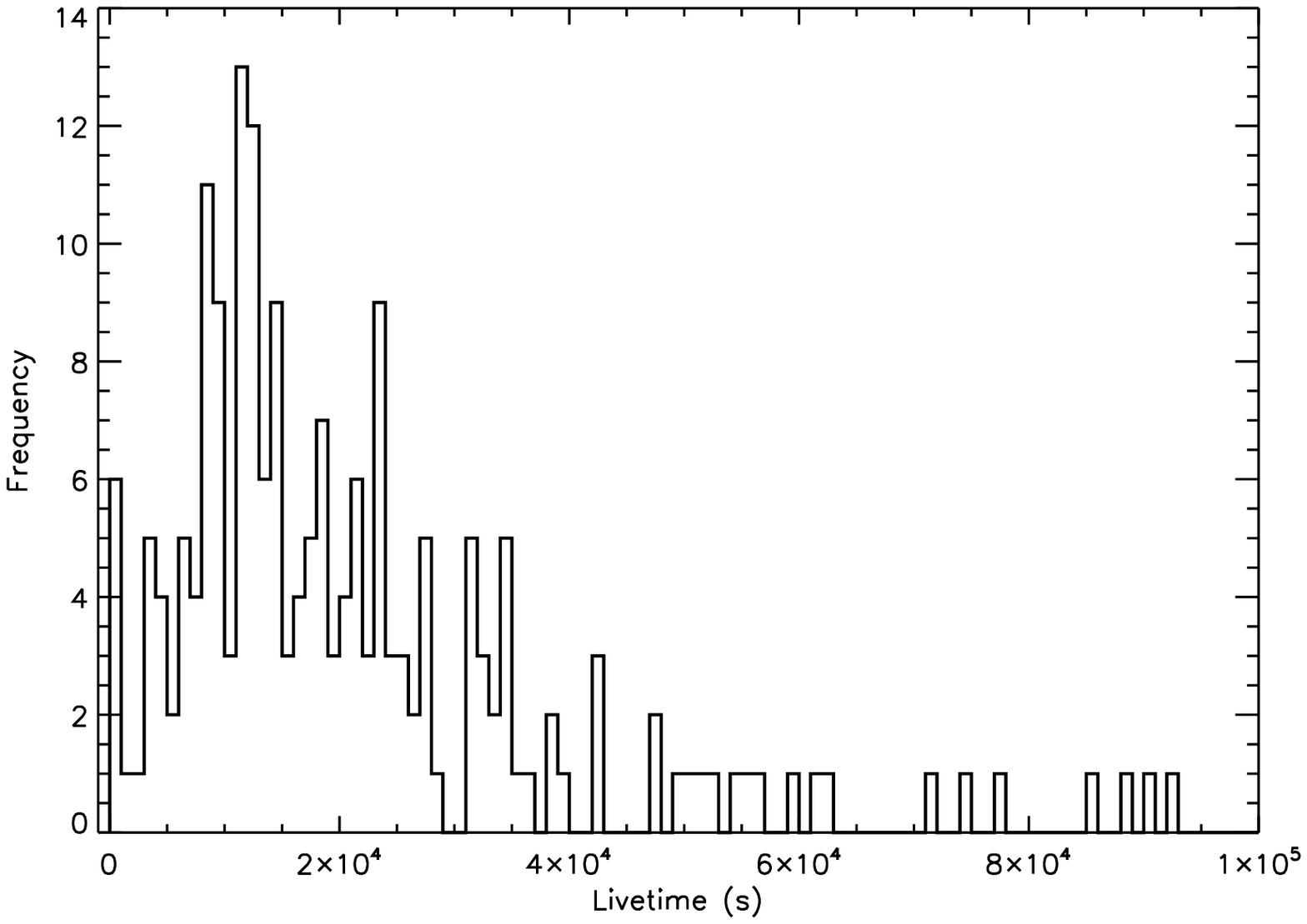}}
     \subfigure[MOS1 revolution number of observation histogram, all observations]{
          \label{revn}
          \includegraphics[width=.45\textwidth]{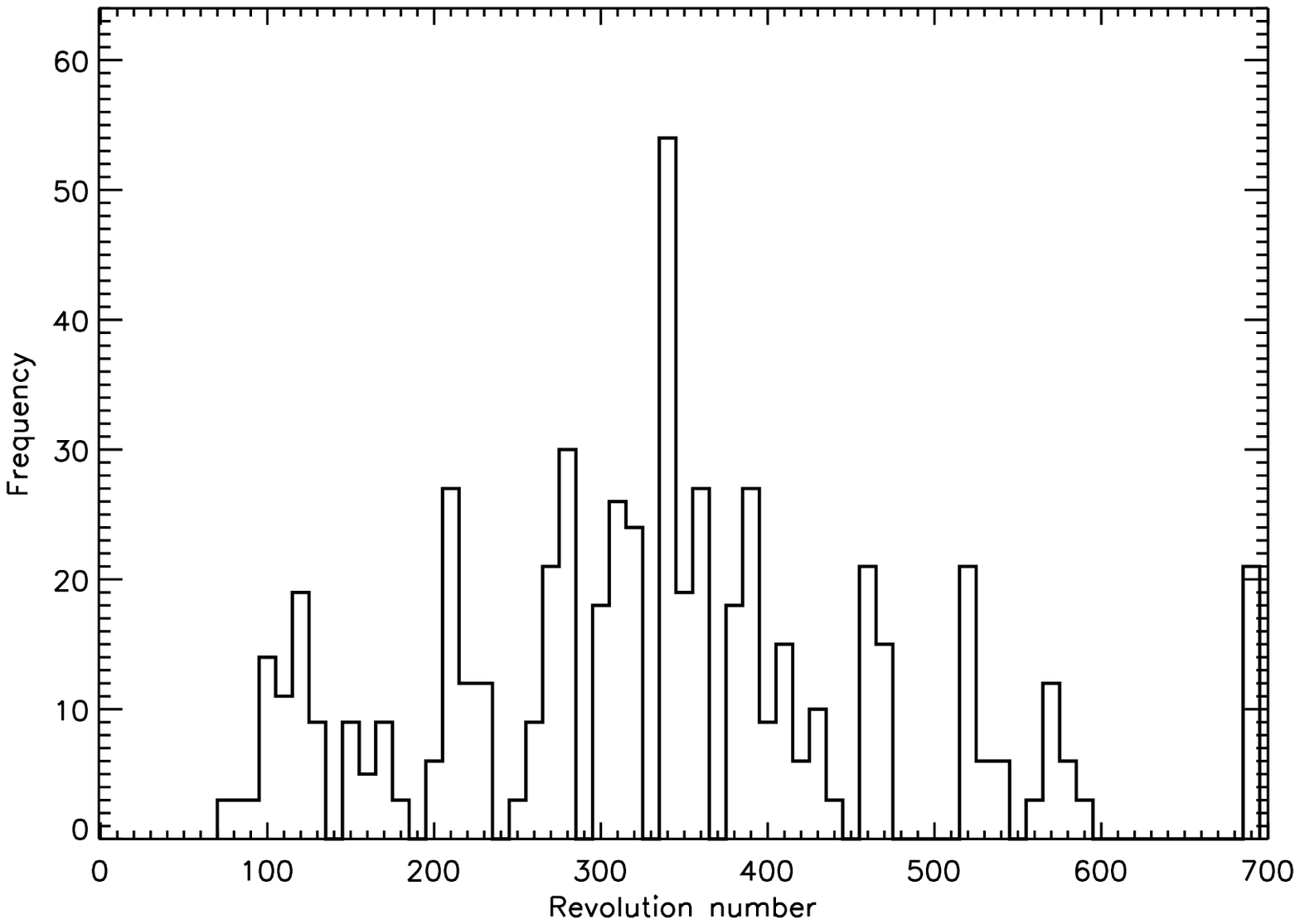}}\\
     \subfigure[MOS1 fraction of time removed, all observations]{
           \label{time}
            \includegraphics[width=.45\textwidth]{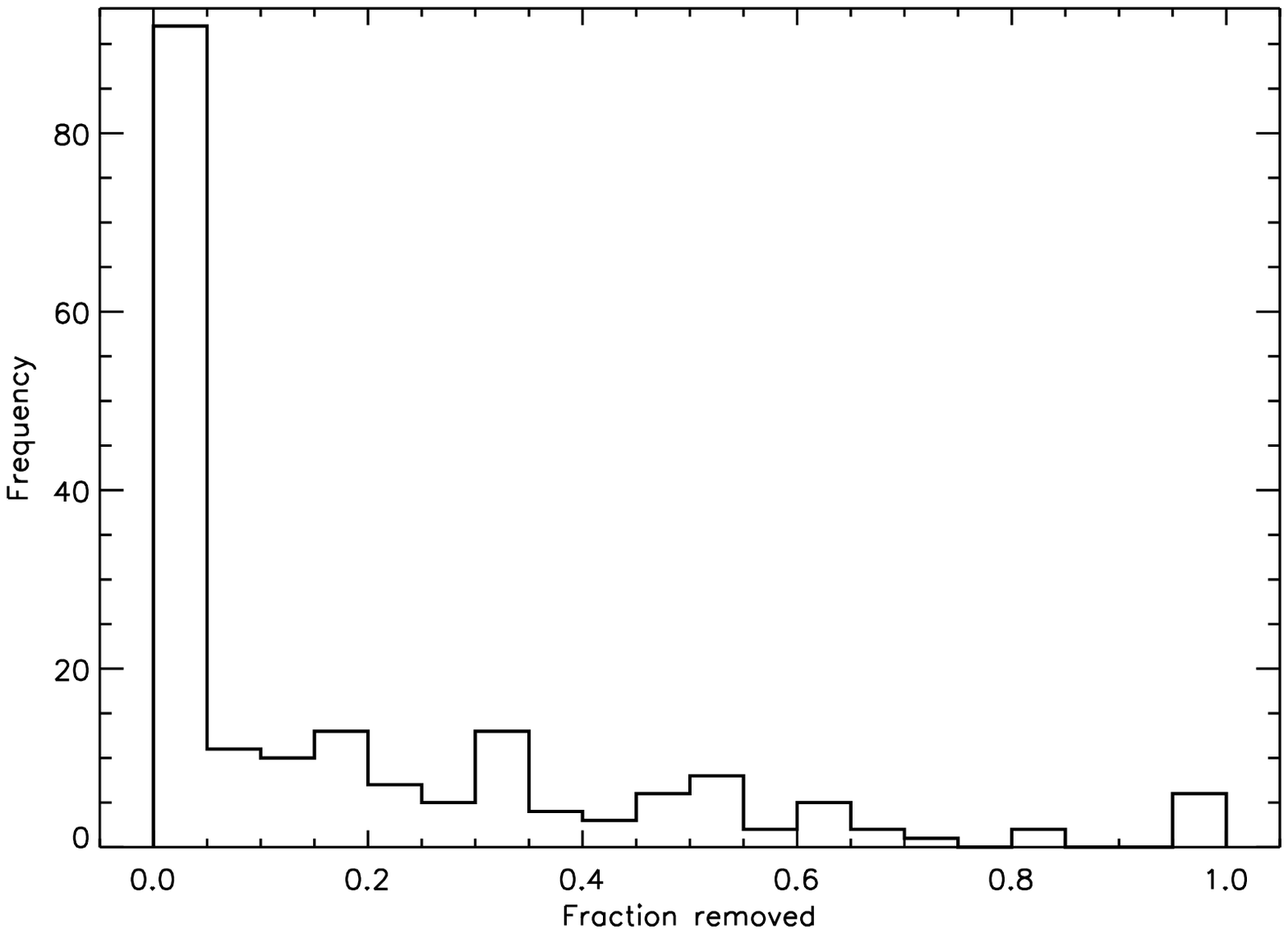}}
     \subfigure[MOS1 fraction of time removed per month, all observations]{
           \label{timemonth}
          \includegraphics[width=.45\textwidth]{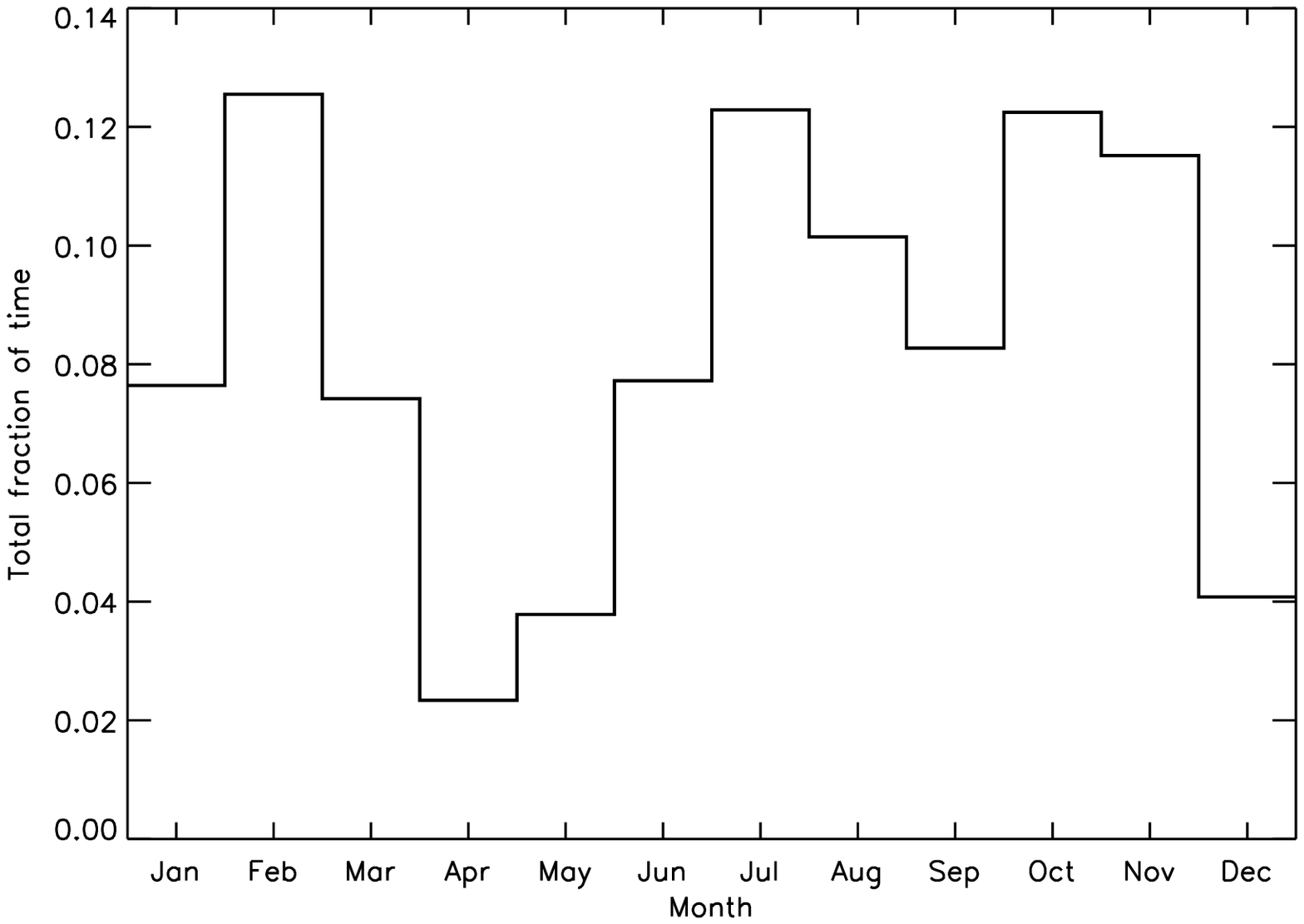}}
     \subfigure[MOS1 histogram of galactic column density values, all observations]{
           \label{nH}
            \includegraphics[width=.45\textwidth]{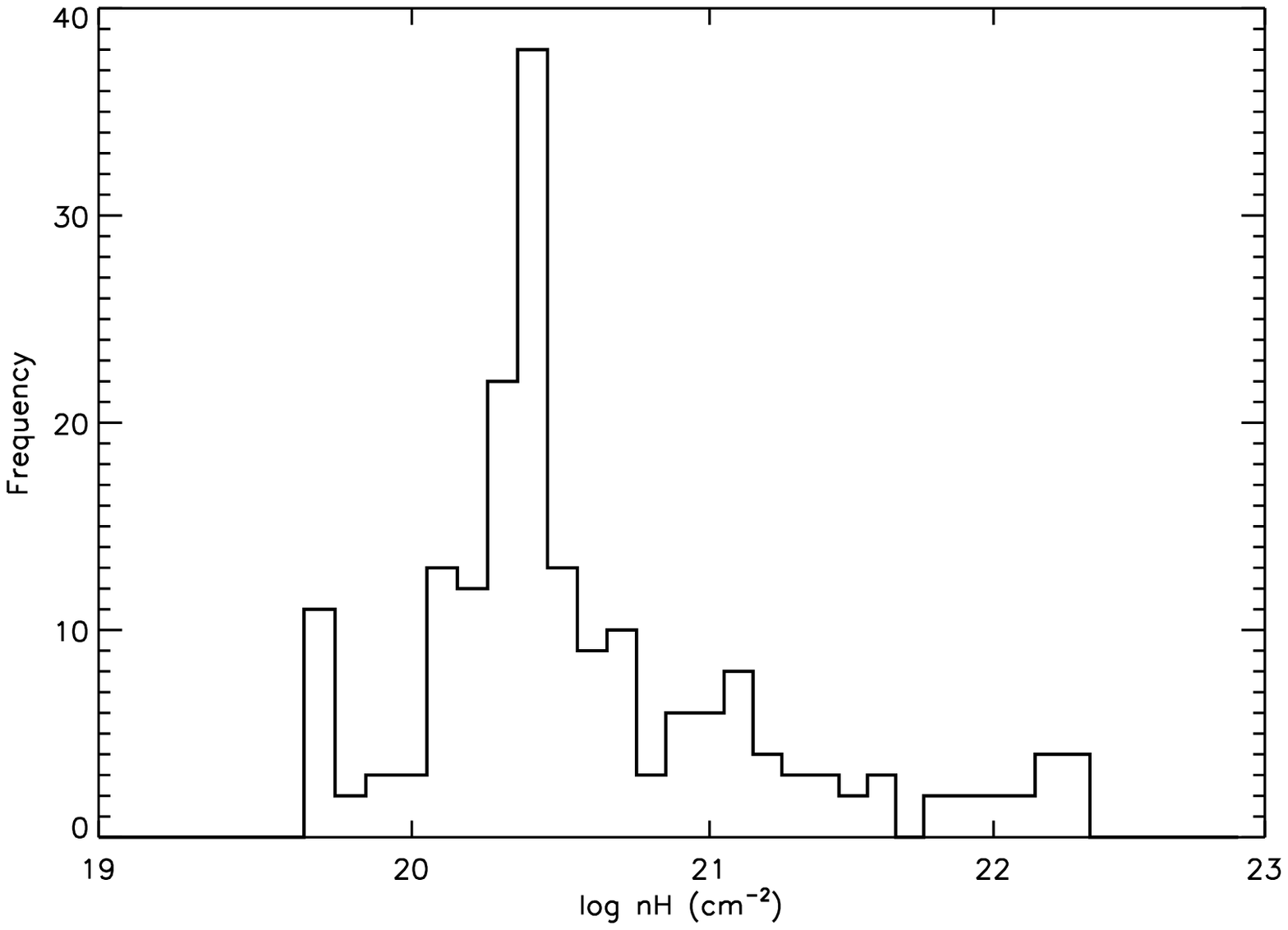}}
     \subfigure[MOS1 livetime per month after cleaning, all observations]{
           \label{livemonth}
          \includegraphics[width=.45\textwidth]{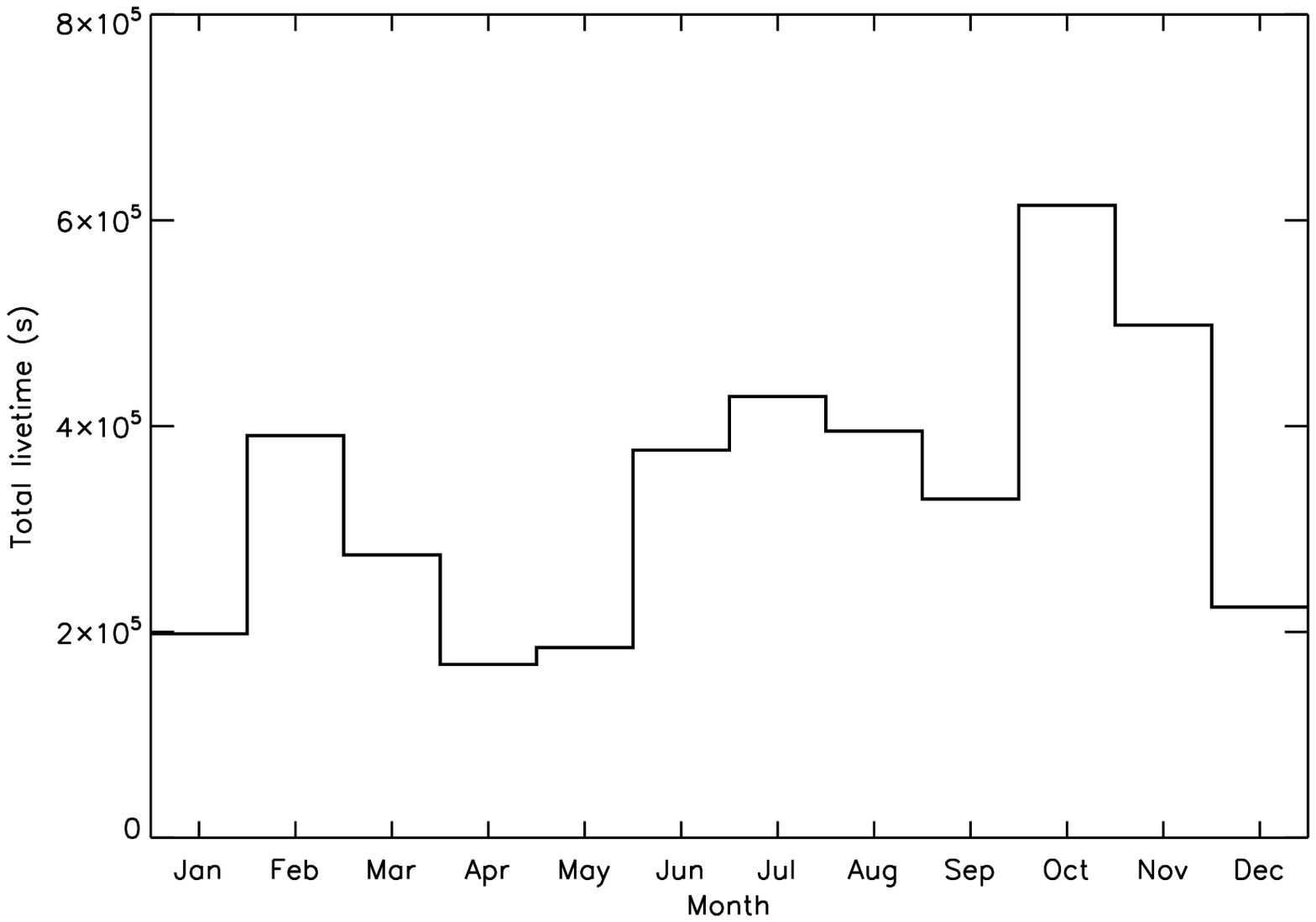}}
     \caption{Properties of all the files used to construct the MOS1
       blank sky event files; (a) histogram of livetimes, (b)
       histogram of revolution numbers, (c) fraction of time removed
       during cleaning, (d) fraction of time removed per month, (e)
       histogram of nH values in the direction of pointing for an
       observation and (f) total livetime after cleaning per month.}

     \label{figobsdetails}
\end{figure*}

Some features of the data sets used for MOS1 can be seen in figure
\ref{figobsdetails}. The equivalent figures for MOS2 and PN look
essentially identical. Figure \ref{liveclean} shows a histogram of the
livetime values for MOS1 after cleaning. The average values of
livetimes after cleaning were 21647 seconds for MOS1, 21682 second for
MOS2 and 20357 seconds for PN, hence the files are very representative
of the XMM-Newton average exposure of $\sim$22950 seconds (based on
$\sim$11200+ EPIC exposures). Observations were taken for all
instruments from between revolution 70 and 691, and hence cover a good
fraction of the mission. The distribution of revolution numbers for
MOS1, is shown in Figure \ref{revn}. This figure also plots a
histogram of the fraction of time removed per observation, and also
the fraction of time removed per month, for MOS1 (\ref{time} and
\ref{timemonth}). The average fraction of time removed for MOS1 was
0.8098, 0.8190 for MOS2 and 0.9577 for PN, which reflects the
conservative cutting levels used in the cleaning.  There is no
evidence that the month of the observation affects the fraction of
time removed from the observation.  Figure \ref{nH} shows a histogram
of the column density values towards each MOS1 pointing. The column
density values have a peak at $\sim$2.5$\times$10$^{20}$ cm$^{-2}$,
which is very typical for an average XMM-Newton pointing. Figure
\ref{livemonth} shows the total livetimes after cleaning for MOS1 per
month. There is a fairly even distribution of observations throughout
the year.

\begin{figure*}[htp]
     \centering
          \includegraphics[width=.3\textwidth]{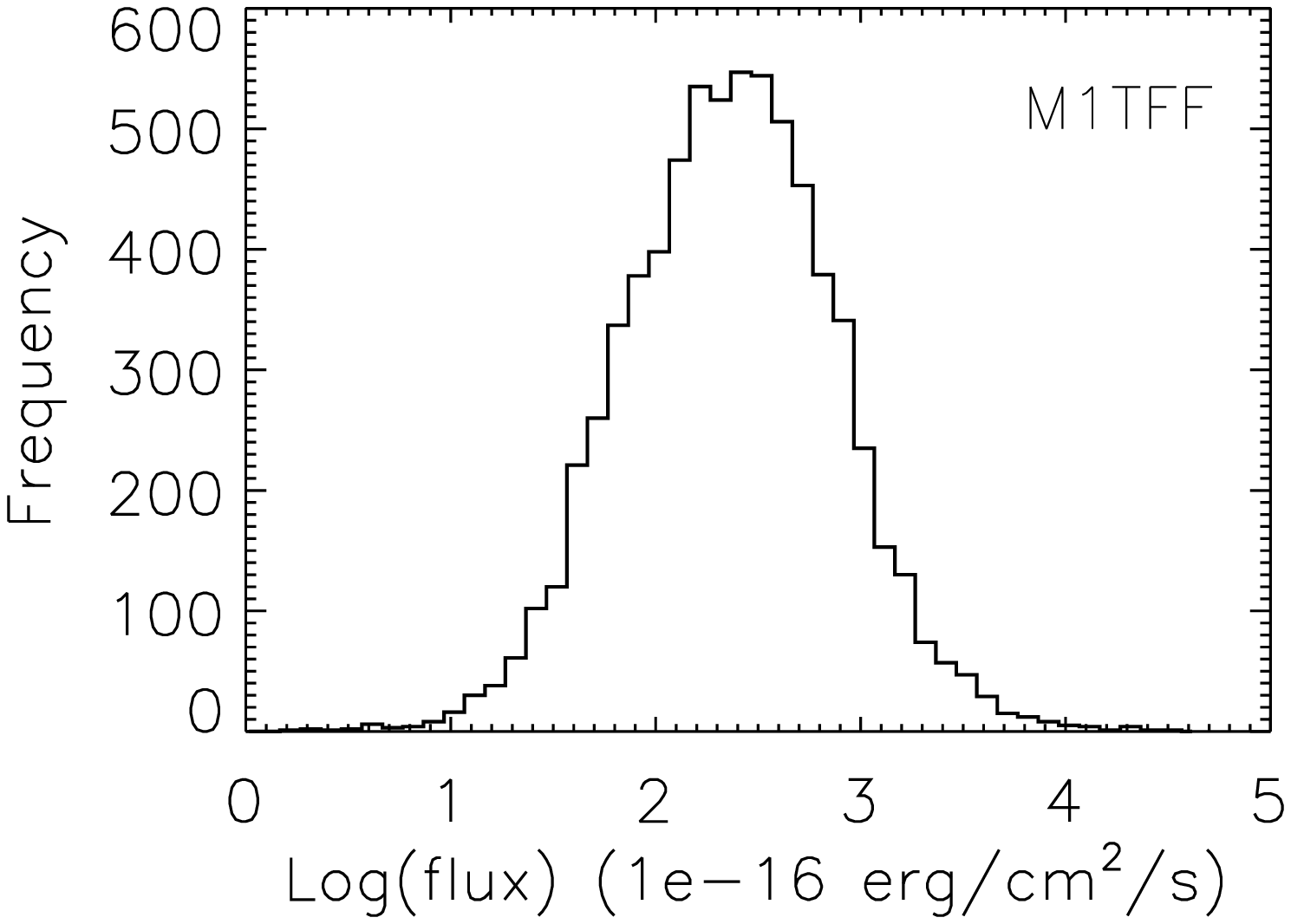}
          \includegraphics[width=.3\textwidth]{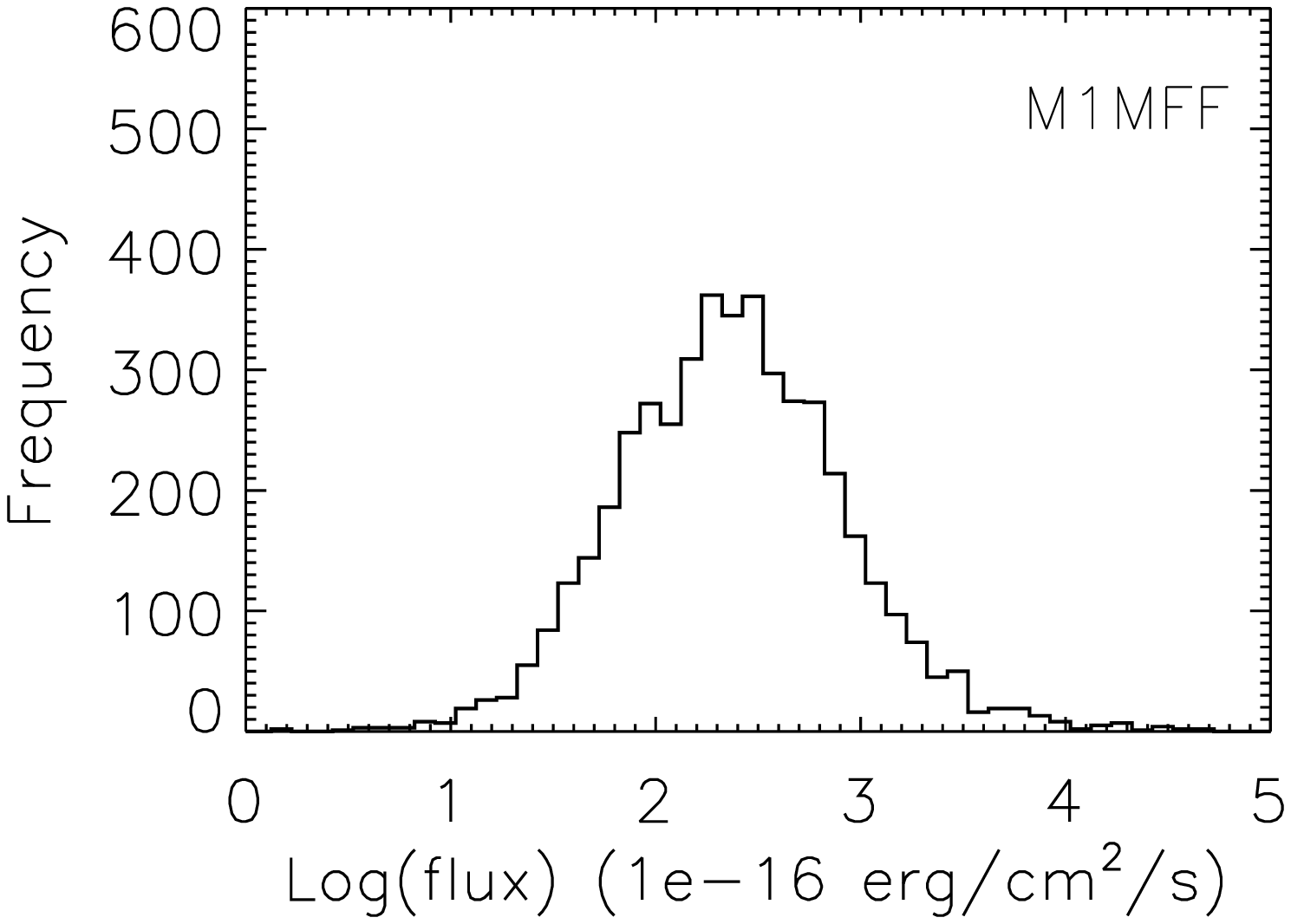}
          \includegraphics[width=.3\textwidth]{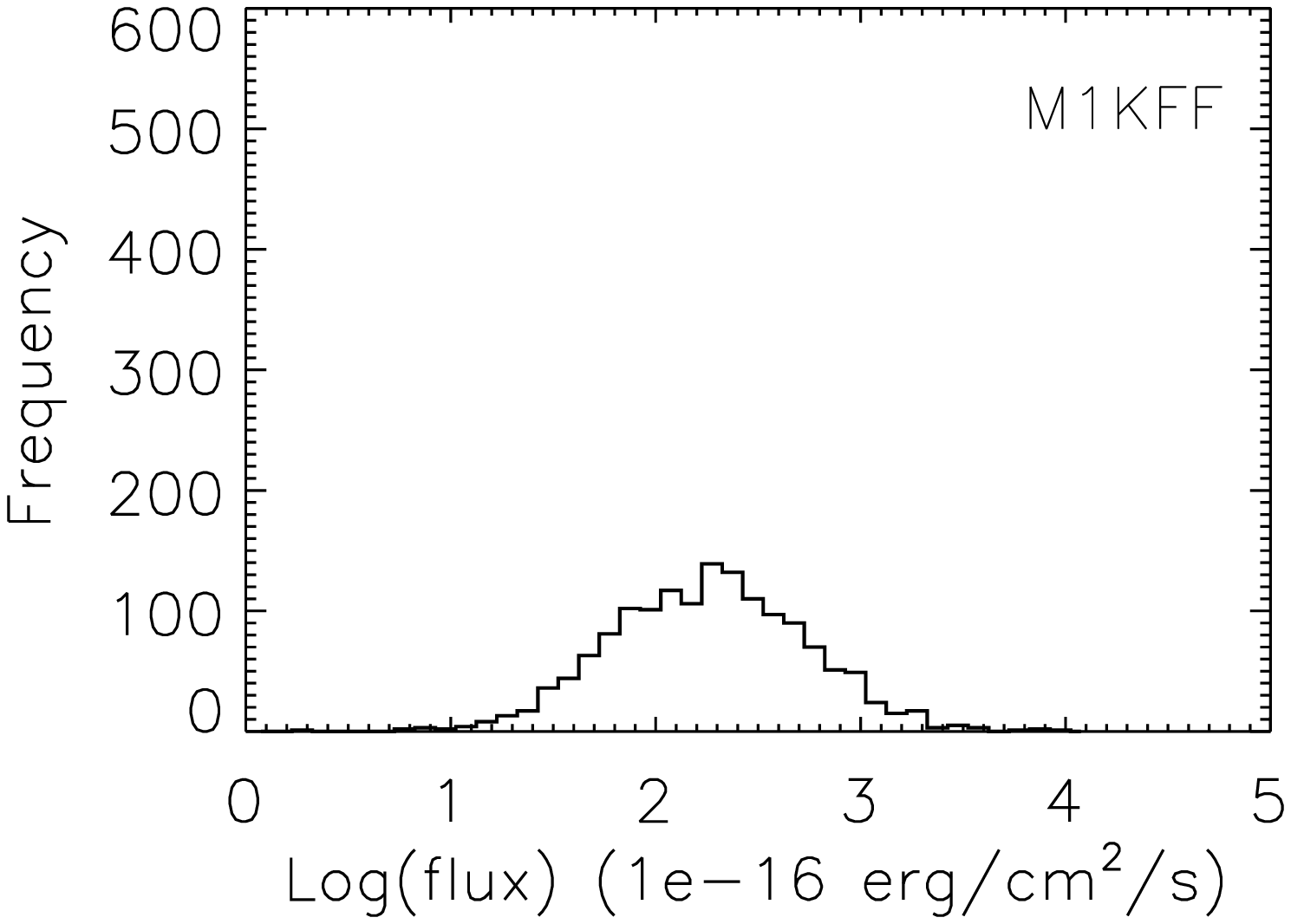}
          \includegraphics[width=.3\textwidth]{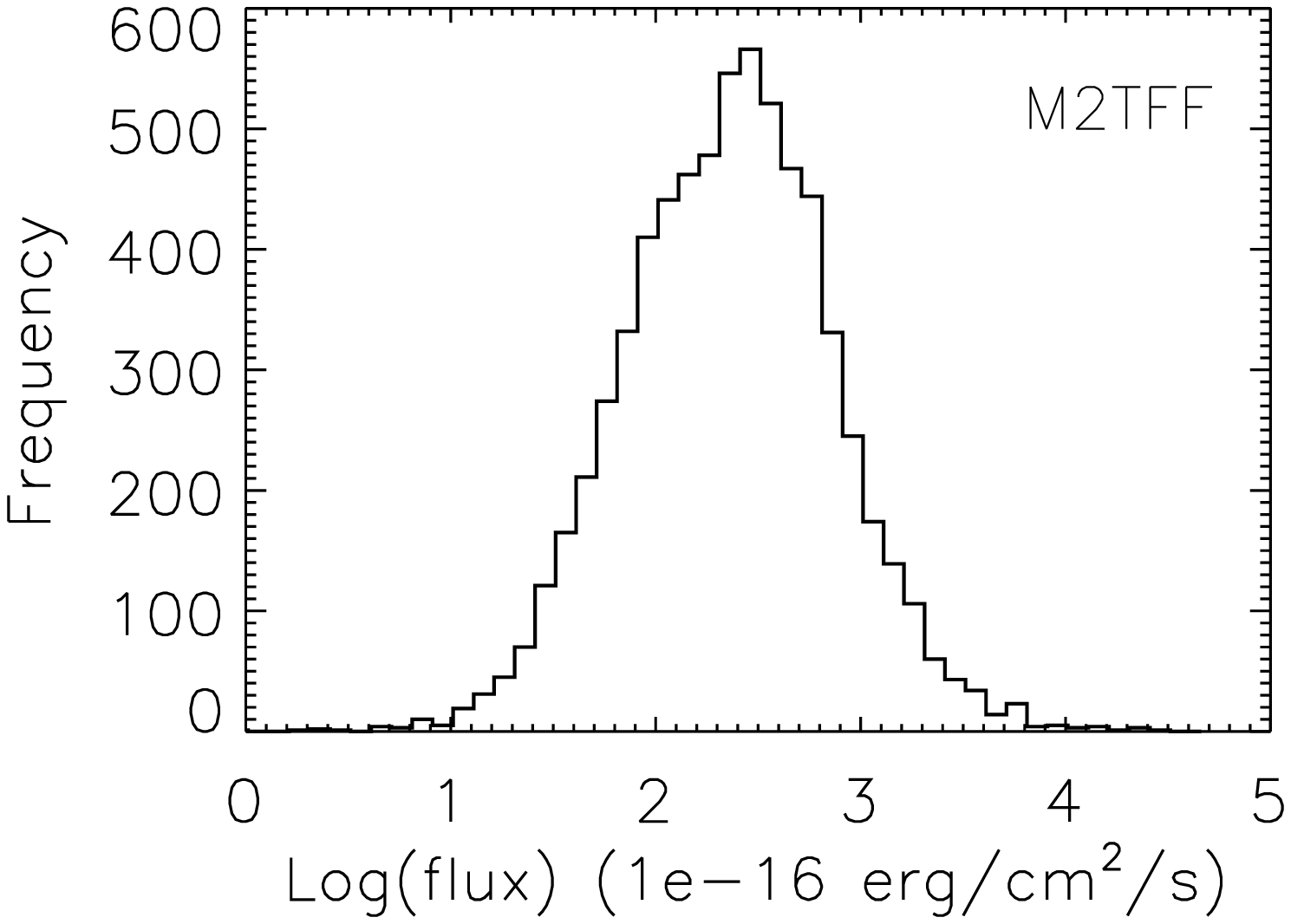}
          \includegraphics[width=.3\textwidth]{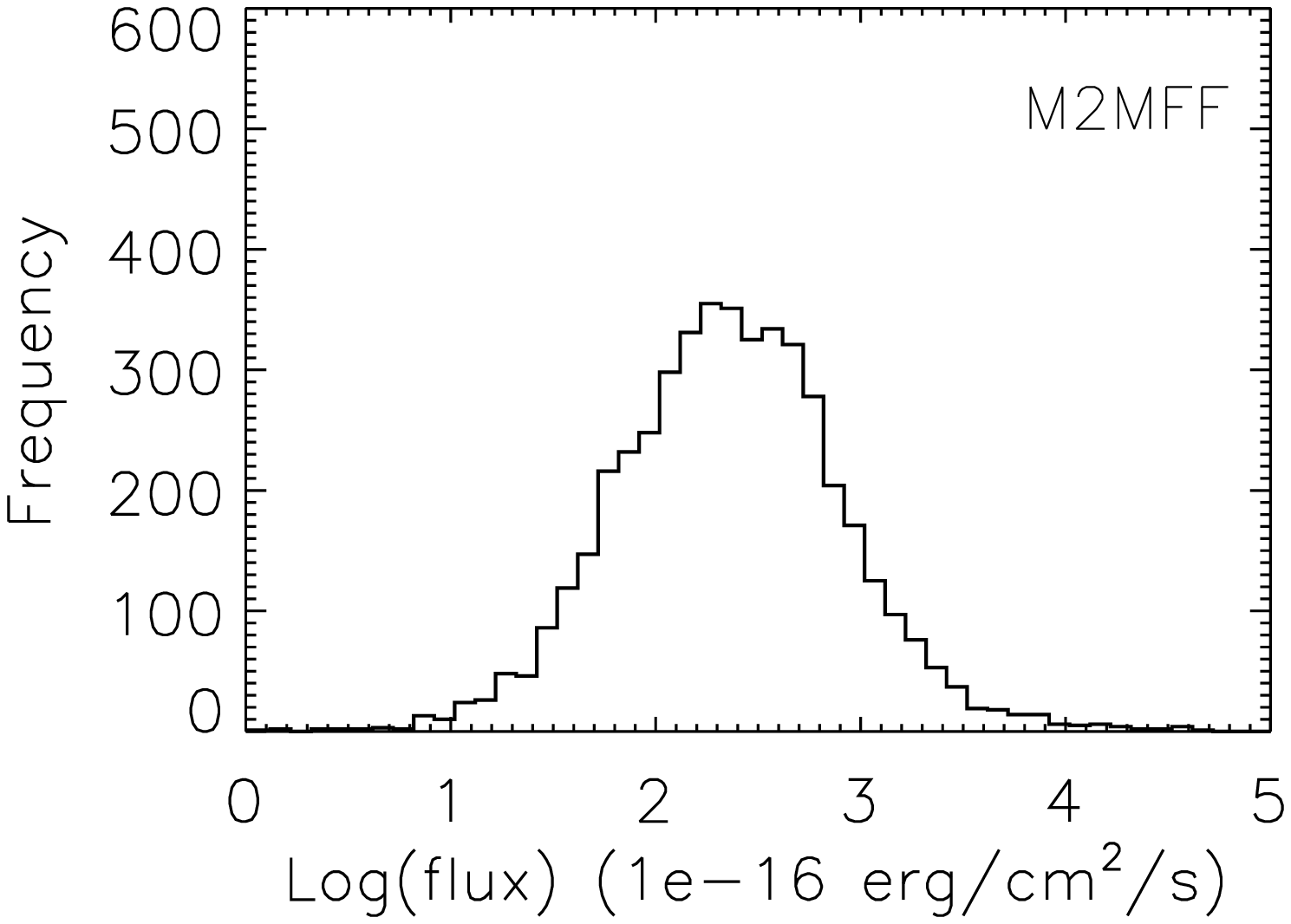}
          \includegraphics[width=.3\textwidth]{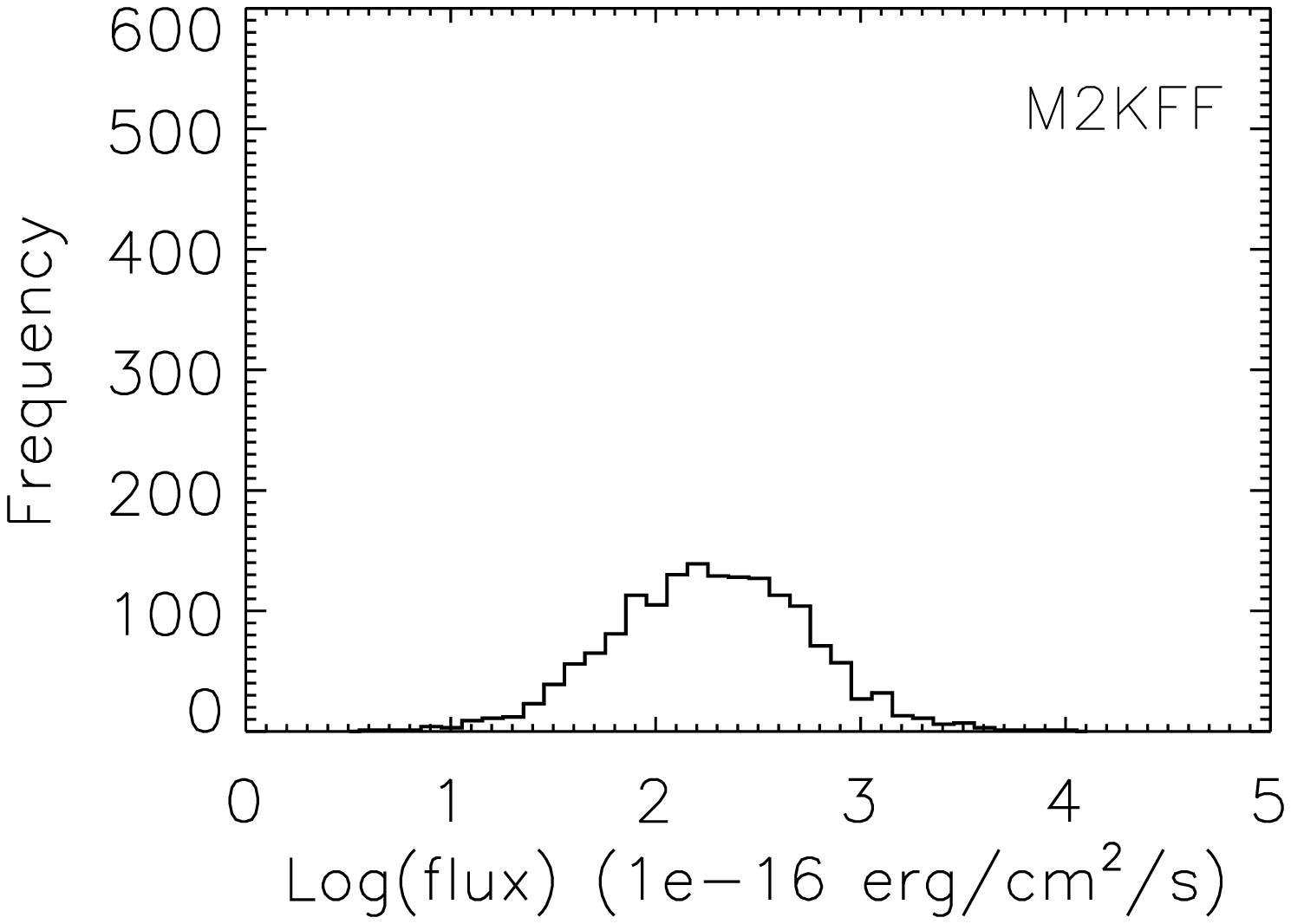}
          \includegraphics[width=.3\textwidth]{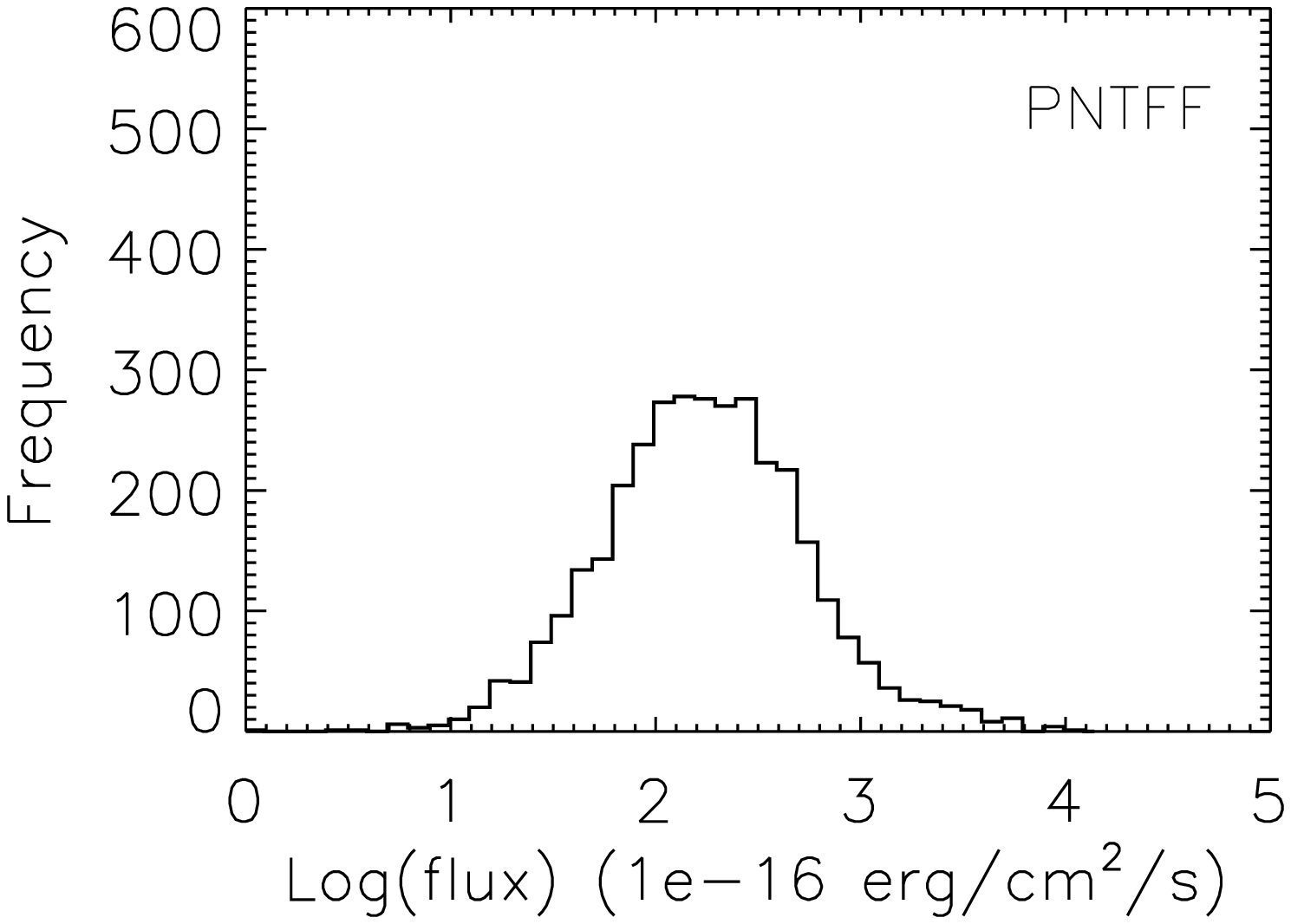}
          \includegraphics[width=.3\textwidth]{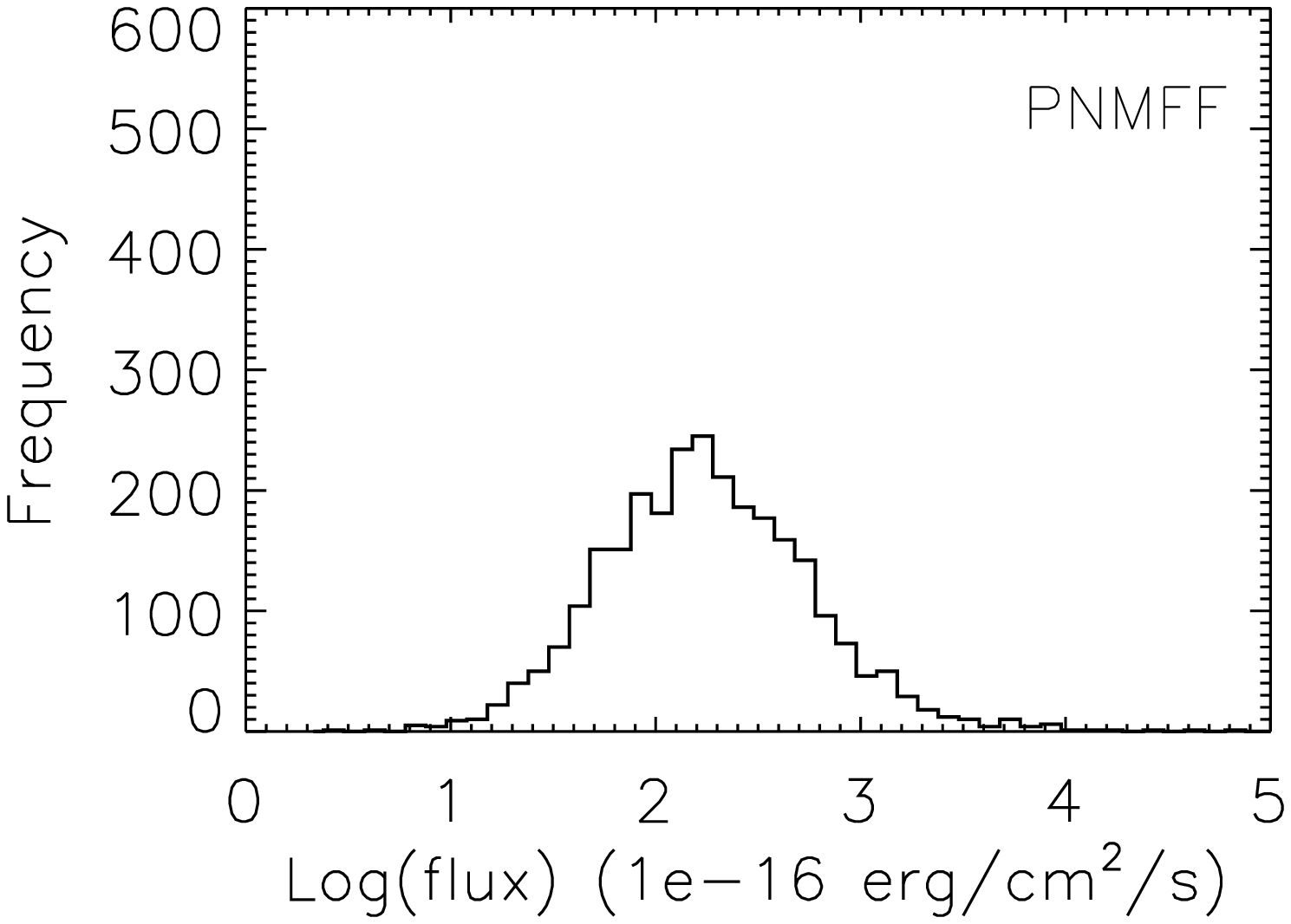}
          \includegraphics[width=.3\textwidth]{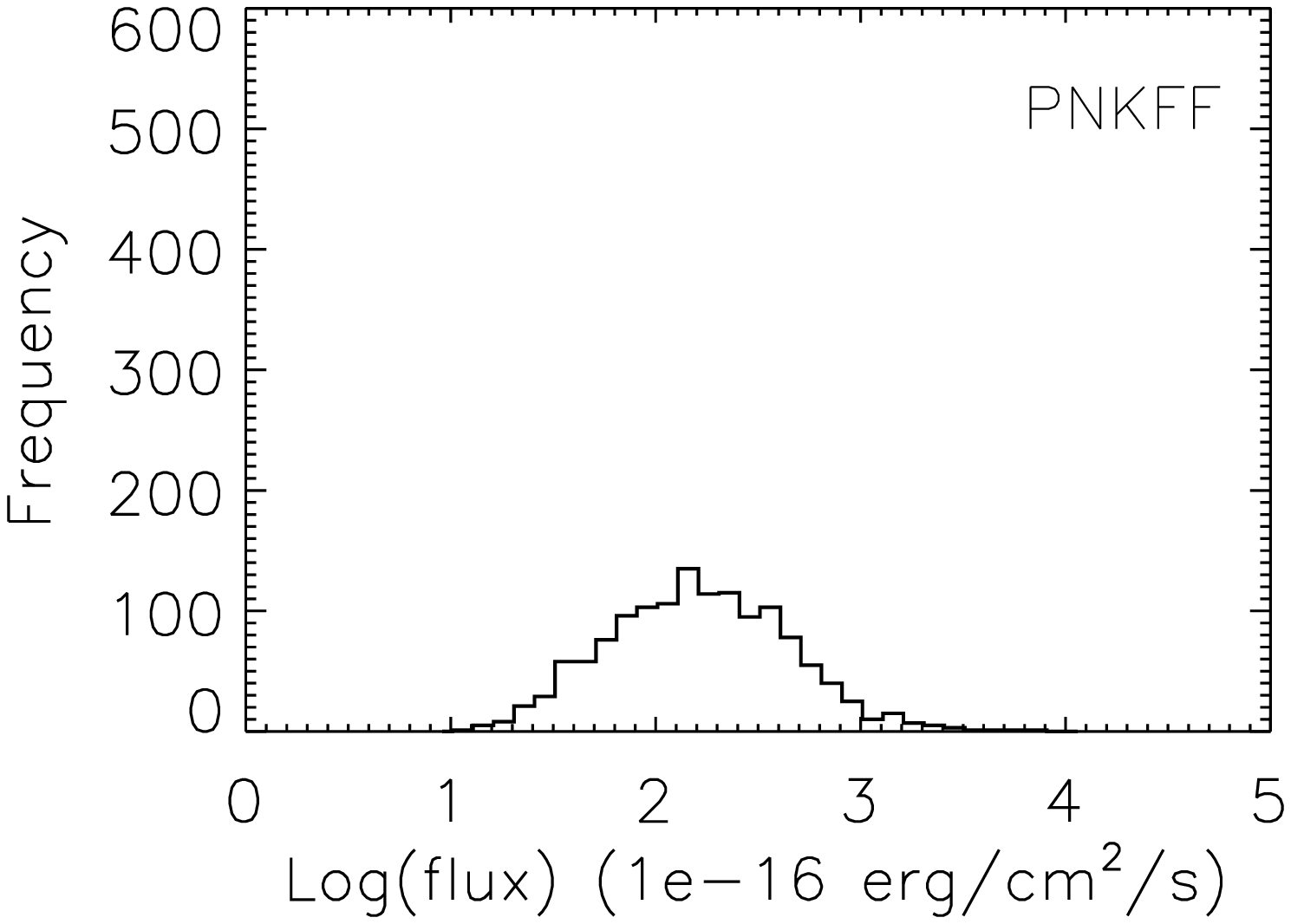}
          \includegraphics[width=.3\textwidth]{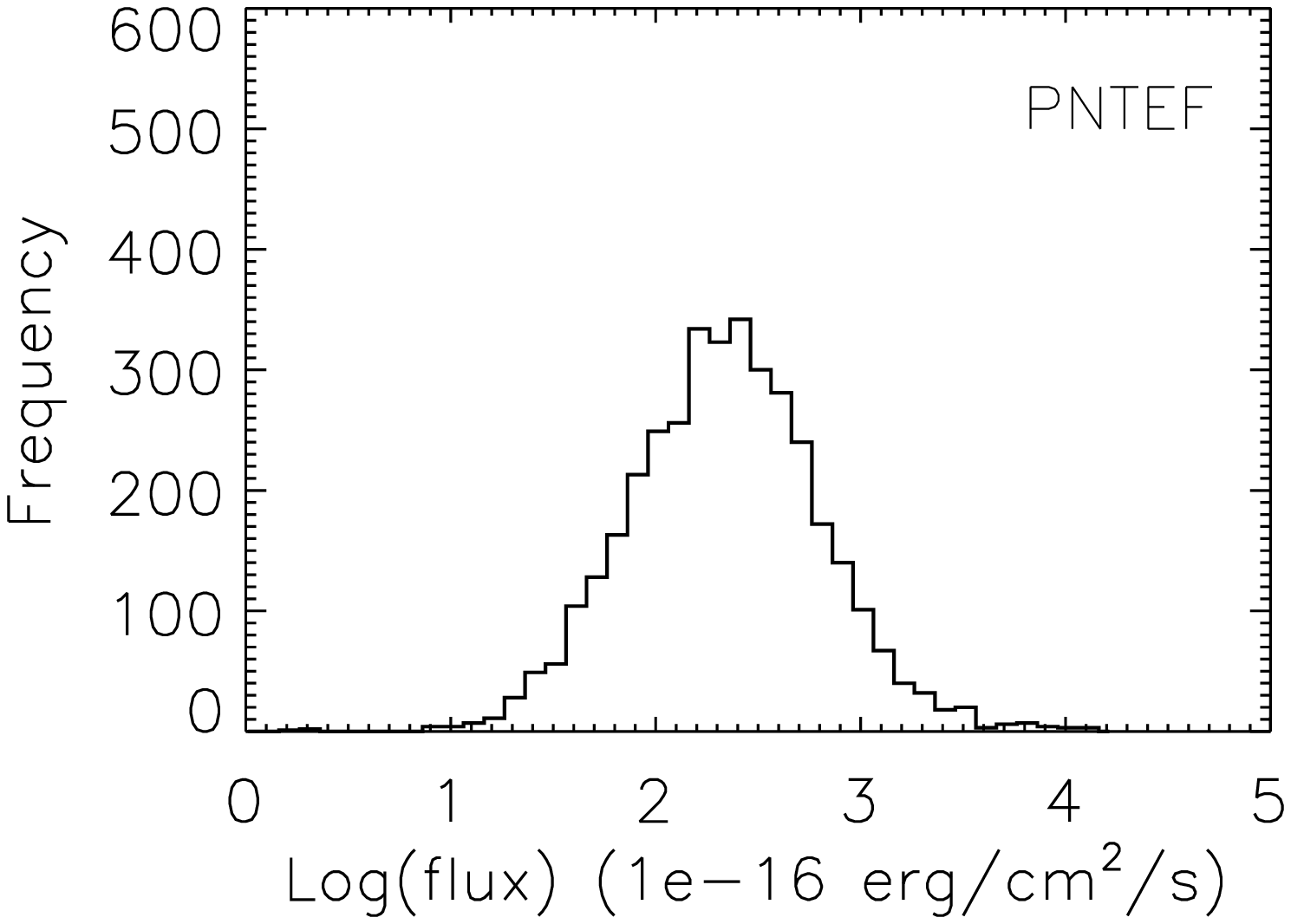}
          \includegraphics[width=.3\textwidth]{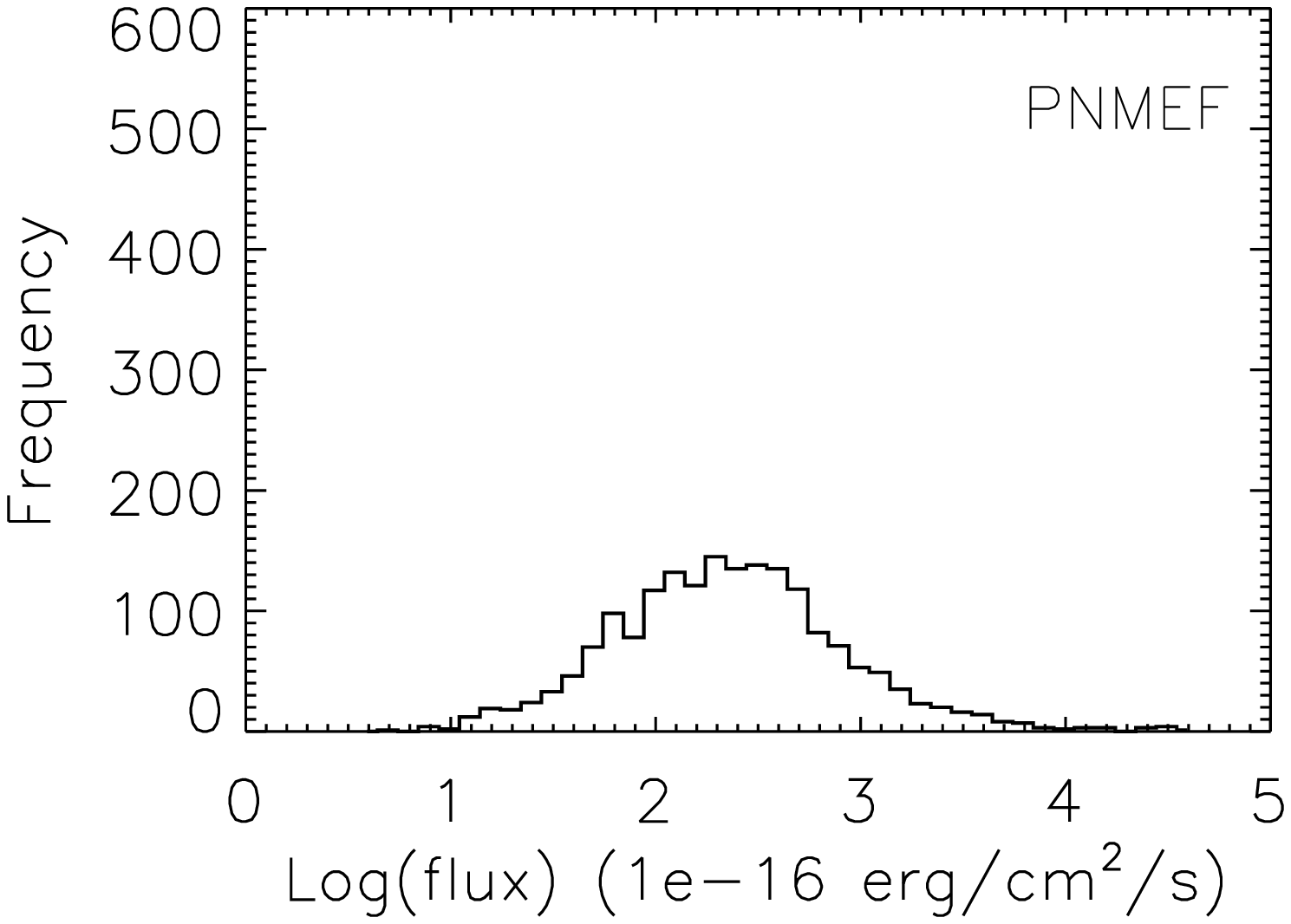}
          \includegraphics[width=.3\textwidth]{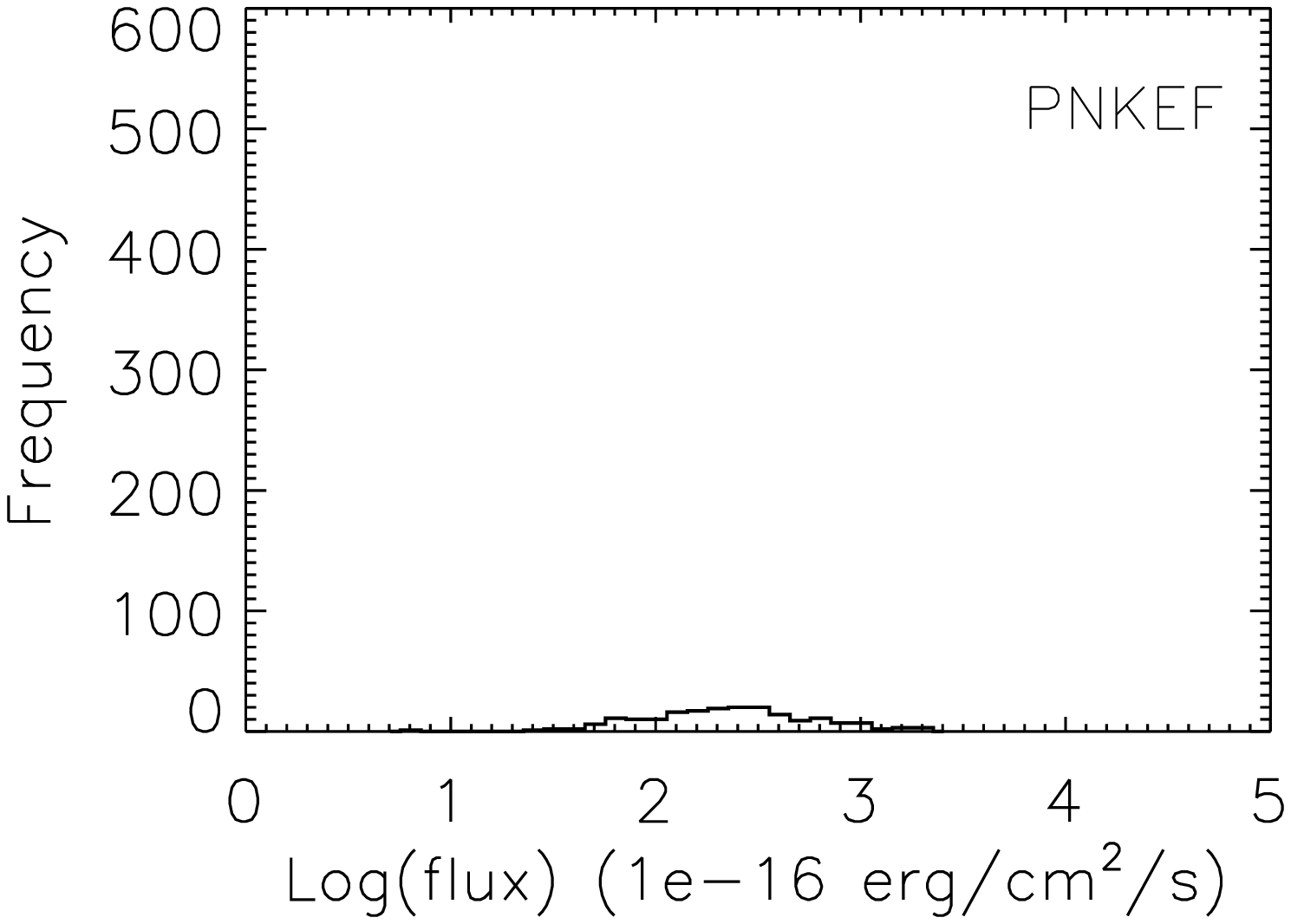}
     \caption{Flux histograms of the removed sources for each of the
instrument-filter-mode combinations data sets, where M1=MOS1, M2=MOS2,
F=full-frame mode, E=full-frame-extended mode, T=thin filter, M=medium
filter and K=thick filter. The y-axis represents the number of sources
removed from the data sets for a particular flux.}
     \label{figfluxremo}
\end{figure*}

Figure \ref{figfluxremo} displays histograms of the fluxes of the
sources removed during the cleaning procedure, for each
instrument-filter-mode combination.  The wide range in observation
times used here has led to the low flux edges being rather shallow,
i.e.\, no particular single flux limits exist above which sources have
been removed. For a general user, this could lead to some data
extracted from the BG event files as being not entirely appropriate
$-$ there possibly being slightly too little of too much emission from
low-flux sources within the extracted BG data. Future plans for this
project (see Section \ref{concl}) include the addition of software
able to select BG events files from the full BG dataset of
observations, based on the component observation's exposure time. This
can lead to BG files with tighter, steeper source flux limit cut-offs,
perhaps more appropriate to a user's own single dataset.

\subsection{Production of the final files}

After the procedure described above was completed for each of the 189
observations, the individual files were reviewed to ensure that no
bright sources remained, relevant event files for each
instrument-filter-mode combination were merged together to form the
final sets (Table \ref{finalfilestable}). Each merged file was
subjected to a call to \textit{attcalc} to set the right ascension,
declination and position angle to (0, 0, 0). In these final files one
will find the primary header, events extension, exposure extensions
and GTI extensions. There are 12 exposure and 12 GTI extensions for PN
and 7 of both types for each MOS.  Keywords that are applicable to
only individual observations, such as \textit{DATE\_OBS} and
\textit{EXPIDSTR} were removed from each file extension.  The keywords
for \textit{LIVETIME}, \textit{LIVETI0n}, \textit{ONTIME}, and
\textit{ONTIME0n} , where n is the number of the CCD, were
recalculated and added to the headers of the final file as
appropriate. All keywords that are mandatory for OGIP FITS
(OGIP\_93\_003) standards were assured to be present and adjusted.
For example the keyword \textit{REVOLUT} was set to 0000,
\textit{OBS\_ID} was set to 0000000000 and \textit{EXP\_ID} was set to
0000000000000.  \\

For both unfilled and refilled PN event files, in the full-frame mode
and either the medium or thin filter, it was found that the resultant
final file exceeded the size limits for use with the XMM-SAS.  It was
necessary, therefore to split these files so that they could be
usable, and a decision was made to split them based on location on the
sky of the events. The files roughly split into hemispheres centred
about the galactic centre and the galactic anti-centre as a variation
between these two locations is seen in spectral shape, as described in
section \ref{propfinal}. We here used the addition of the right
ascension and declination (\textit{RA} and \textit{DEC}) columns added
to the component event files in the individual observation processing,
which then later appear in the final event files.\\

For both the unfilled or refilled event lists, this has resulted in
the production of 14 final background event files. For each event file
there is a corresponding non-vignetted and vignetted exposure
map. Therefore there are 84 files in total (42 for either unfilled or
refilled procedures).

Table \ref{finalfilestable} summarises the resulting background event
files, detailing the number of component observations used and the
overall exposure time.

\begin{figure*}
\vspace{1cm}
\includegraphics[height=20cm, width=16cm]{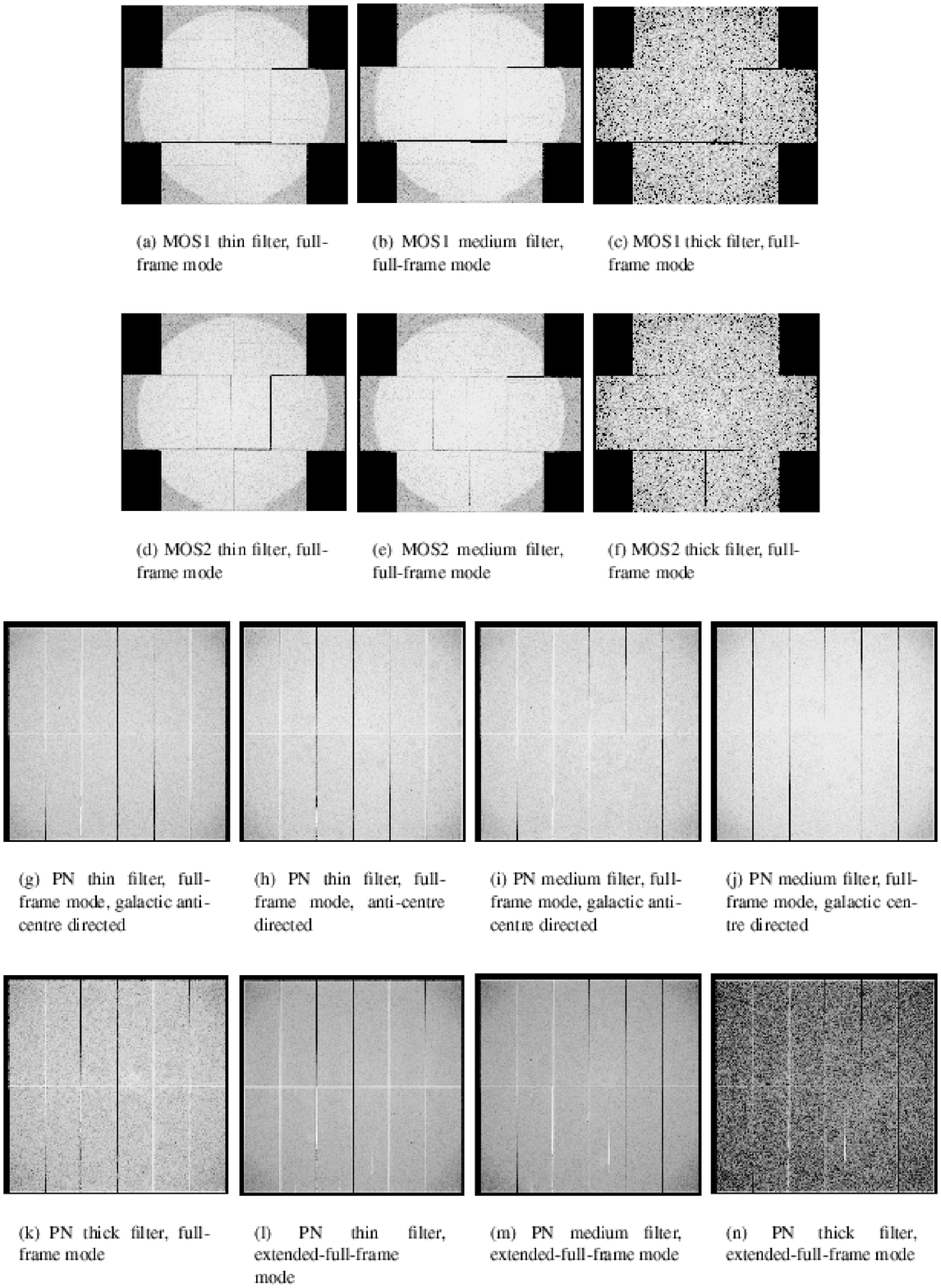}
     \caption{Images of the final unfilled event files for each
       instrument-filter-mode combination, between 0.3\,keV and 12\,keV
       for the PN and from between 0.2\,keV and 12\,keV for the MOS
       cameras. The PN event files for full-frame mode, with the
       medium or thin filters have been split into those events
       centred about the galactic centre, and those centred about the
       galactic anti-centre.}
\label{figfinalfiles}
\end{figure*}

These final event files offer several improvements over previous blank
sky background event files. The ghosting procedure is an addition to
the previous work and we have taken advantage of the new 2XMMp
reprocessing products and improved general knowledge of the BG to
improve the cleaning of each file. These files are found at: \newline
\textit{http://xmm.vilspa.esa.es/external/xmm\_sw\_cal} \newline
\indent \textit{/background/index.shtml}. \newline The site includes a
description of the naming convention of these files. Images created
from the final filled event files can be seen in Figure
\ref{figfinalfiles}.

\begin{table*}
\caption[]{{\small Summary of all the observations used. IMF gives the
instrument, mode and filter for PN-MOS1-MOS2 as [f=full-frame,
e=full-frame-extended, t=thin, m=medium, k=thick and NA - not
applicable]. Exp is the cleaned exposure time, and f(Ex) the fraction
of exposure removed after screening (where possible PN or
alternatively MOS1).}}  {\tiny{
\begin{tabular}{|cccc|cccc|cccc|}
\hline
Obs & IMF   & Exp.(s)  & f(Ex) & Obs & IMF & Exp. & f(Ex) & Obs & IMF & Exp.(s) & f(Ex)  \\ 
\hline
0123700101 & ftftfk & 46561 & 0.192 & 0125300101 & ftftft & 38127 & 0.002 & 0126511201 & fkfkfk & 29205 & 0.018 \\
0110980401 & etftft & 39999 & 0.015 & 0112940201 & etftft & 5898 & 0.002 & 0112940301 & etftft & 6099 & 0.000 \\
0112940401 & etftft & 5898 & 0.000 & 0112940501 & NAftft & 7184 & 0.254 & 0127921001 & etftft & 51496 & 0.000 \\
0127921201 & etftft & 14699 & 0.000 & 0112370101 & NAftft & 57523 & 0.168 & 0112371001 & ftftft & 56766 & 0.075 \\
0102640201 & emfmfm & 13300 & 0.000 & 0112370301 & NAftft & 63066 & 0.204 & 0112371501 & ftftft & 8300 & 0.019 \\
0112371701 & ftftft & 25247 & 0.002 & 0129320801 & emfmfm & 6199 & 0.000 & 0129320901 & etftft & 6197 & 0.000 \\
0129321001 & NAftft & 5039 & 0.000 & 0112230901 & etftft & 23948 & 0.307 & 0112231001 & etftft & 24485 & 0.004 \\
0112970701 & emfmfm & 19519 & 0.000 & 0094800201 & etftft & 48056 & 0.306 & 0110660301 & etftft & 4855 & 0.110 \\
0104460301 & fmfmfm & 9649 & 0.000 & 0128530301 & NAftft & 37446 & 0.153 & 0113891201 & fmfmfm & 19456 & 0.052 \\
0106660101 & ftftft & 54807 & 0.000 & 0099030101 & ftNANA & 19946 & 0.048 & 0110660401 & emfmfm & 9287 & 0.015 \\
0110980101 & etftft & 50256 & 0.046 & 0112650401 & etftfm & 19999 & 0.000 & 0134531201 & fkfkfk & 18999 & 0.000 \\
0134531301 & fkfkfk & 6024 & 0.000 & 0007420701 & fmfmfm & 9700 & 0.000 & 0007420801 & fmfmfm & 12305 & 0.000 \\
0007421001 & fmfmfm & 9514 & 0.048 & 0007421901 & fmfmfm & 7999 & 0.000 & 0007422001 & fmfmfm & 7599 & 0.000 \\
0007422101 & fmfmfm & 9400 & 0.000 & 0007422201 & fmfmfm & 9400 & 0.000 & 0007422301 & fmfmfm & 12650 & 0.000 \\
0093640901 & emfmfm & 5899 & 0.007 & 0111120201 & fmfmfm & 30377 & 0.000 & 0022140101 & emfmfm & 11156 & 0.000 \\
0093670501 & emfmfm & 9155 & 0.000 & 0021750701 & fkfkfk & 25649 & 0.000 & 0024140101 & fkfkfk & 58270 & 0.000 \\
0067340201 & emfmfm & 9949 & 0.000 & 0135740901 & fmfmfm & 9250 & 0.000 & 0081341001 & ftftft & 19602 & 0.060 \\
0006010301 & fkfkfk & 31929 & 0.000 & 0070340301 & emfmfm & 26429 & 0.473 & 0111550101 & ftftft & 42381 & 0.013 \\
0111550201 & ftftft & 41095 & 0.012 & 0092970201 & emfmfm & 8980 & 0.000 & 0093640701 & emfmfm & 15065 & 0.000 \\
0111550401 & ftftft & 91805 & 0.005 & 0134531401 & fkfkfk & 11842 & 0.709 & 0032140201 & fmfmfm & 9839 & 0.135 \\
0109660801 & ftftfm & 62945 & 0.076 & 0051760201 & ftftft & 18260 & 0.047 & 0109660901 & ftfmft & 42812 & 0.341 \\
0109661001 & ftftft & 80575 & 0.077 & 0110990201 & etftft & 24614 & 0.189 & 0109520501 & etftft & 19960 & 0.000 \\
0111110301 & etftft & 20010 & 0.000 & 0111110401 & etftft & 22421 & 0.000 & 0111110501 & etftft & 20039 & 0.000 \\
0111110101 & etftft & 20898 & 0.009 & 0111110201 & etftft & 8196 & 0.263 & 0111280301 & etftft & 4960 & 0.119 \\
0112250301 & etftft & 21955 & 0.000 & 0137750101 & ftftft & 16360 & 0.000 & 0057560301 & ftftft & 37681 & 0.015 \\
0112220101 & fmfmfm & 37390 & 0.000 & 0093630101 & fmfmfm & 13199 & 0.000 & 0111110701 & etftft & 8399 & 0.000 \\
0000110101 & fmfmfm & 28811 & 0.100 & 0067750101 & ftftft & 44219 & 0.180 & 0016140101 & NAfmfm & 32945 & 0.459 \\
0112190401 & emfmfm & 9994 & 0.002 & 0112410101 & ftftft & 9609 & 0.006 & 0112460201 & fmfmfm & 36447 & 0.135 \\
0002740101 & ftftft & 29685 & 0.102 & 0089940201 & ftftft & 30082 & 0.074 & 0111160201 & etftft & 44373 & 0.001 \\
0112480101 & etftft & 15664 & 0.000 & 0112480201 & etftfm & 13288 & 0.000 & 0094400101 & fmfmfm & 29283 & 0.000 \\
0050940301 & emfmfm & 9098 & 0.035 & 0137950201 & ftftft & 21529 & 0.070 & 0137950301 & ftftft & 22500 & 0.000 \\
0050150301 & ftftft & 25133 & 0.060 & 0109260201 & ekfkfk & 28453 & 0.000 & 0032140501 & fmfmfm & 9971 & 0.000 \\
0092360201 & etftft & 9935 & 0.000 & 0079570201 & emfmfm & 43075 & 0.092 & 0094400301 & fmfmfm & 19978 & 0.000 \\
0021740101 & emfmfm & 29888 & 0.021 & 0022740201 & fmfmfm & 61189 & 0.012 & 0094530401 & etftft & 19923 & 0.000 \\
0092360101 & etftft & 7399 & 0.000 & 0092360301 & etftft & 9872 & 0.110 & 0092360401 & etftft & 8599 & 0.000 \\
0092800101 & fmfmfm & 14277 & 0.000 & 0134531501 & fkfkfk & 18484 & 0.000 & 0109463501 & ftftft & 5000 & 0.000 \\
0028740301 & ftftft & 25863 & 0.000 & 0046940401 & emfmfm & 10598 & 0.000 & 0083240201 & etftft & 15699 & 0.001 \\
0086750101 & ftftft & 8595 & 0.123 & 0022740301 & fmfmfm & 36366 & 0.009 & 0085170101 & ftftft & 29937 & 0.000 \\
0085640201 & etfmfm & 29904 & 0.000 & 0049340201 & fmfmfm & 24558 & 0.000 & 0112190601 & emfmfm & 10540 & 0.000 \\
0106660401 & NAftft & 34391 & 0.025 & 0106660501 & ftftft & 8076 & 0.169 & 0070940101 & NAftft & 9924 & 0.411 \\
0026340201 & emfmfm & 13906 & 0.237 & 0112551201 & emftft & 9999 & 0.000 & 0055990301 & etftft & 9999 & 0.000 \\
0033541001 & ftftft & 9912 & 0.000 & 0052140201 & emfmfm & 36228 & 0.036 & 0059800101 & etftft & 5769 & 0.000 \\
0112550501 & emftft & 18499 & 0.000 & 0011830201 & ftftft & 32653 & 0.065 & 0002970401 & NANAft & 3640 & 0.001 \\
0112810201 & ftftft & 13000 & 0.085 & 0033540901 & ftftft & 14444 & 0.027 & 0037980101 & etftft & 10177 & 0.000 \\
0037980201 & etftft & 8999 & 0.000 & 0037980301 & etftft & 9099 & 0.000 & 0037980601 & etftft & 8992 & 0.000 \\
0037980701 & etftft & 7999 & 0.000 & 0037980401 & etftft & 10378 & 0.071 & 0037980501 & etftft & 12951 & 0.022 \\
0037980901 & etftft & 9999 & 0.000 & 0109520601 & etftft & 18981 & 0.000 & 0109520301 & etftft & 17961 & 0.000 \\
0112680401 & ftftft & 21991 & 0.000 & 0109110101 & emfmfm & 71843 & 0.000 & 0025740401 & ftftft & 17449 & 0.146 \\
0083950101 & emfmfm & 23222 & 0.026 & 0081341101 & ftftft & 17105 & 0.000 & 0109280101 & ekfkfk & 19796 & 0.000 \\
0112480301 & etftft & 13533 & 0.000 & 0050940101 & emfmfm & 19999 & 0.048 & 0006010401 & fkfkfk & 32932 & 0.000 \\
0026340101 & etftft & 21932 & 0.000 & 0058940101 & emfmfm & 23793 & 0.000 & 0110661601 & etftft & 5999 & 0.000 \\
0051610101 & emfmfm & 17960 & 0.000 & 0026340301 & emfmfm & 19829 & 0.008 & 0092800201 & fmfmfm & 92855 & 0.095 \\
0082140301 & emfmfm & 28892 & 0.000 & 0112190201 & emfmfm & 9999 & 0.000 & 0111280601 & etftft & 7176 & 0.000 \\
0111282001 & etftft & 4999 & 0.003 & 0103060301 & etftft & 42728 & 0.034 & 0111280701 & etftft & 4992 & 0.005 \\
0111280801 & etftft & 4999 & 0.000 & 0111282201 & etftft & 4999 & 0.000 & 0037980801 & etftft & 9999 & 0.011 \\
0112551501 & ekftft & 17749 & 0.000 & 0037981001 & etftft & 9999 & 0.000 & 0037981101 & etftft & 9867 & 0.000 \\
0037981201 & etftft & 9753 & 0.045 & 0110662701 & emfmfm & 6291 & 0.383 & 0147510101 & fmfmfm & 90984 & 0.020 \\
0147510801 & fmfmfm & 75660 & 0.047 & 0147510901 & fmfmfm & 88534 & 0.066 & 0147511101 & fmfmfm & 93394 & 0.216 \\
0147511201 & fmfmfm & 100499 & 0.363 & 0147511301 & fmfmfm & 82459 & 0.182 & 0145450401 & ftftft & 5748 & 0.000 \\
0145450501 & ftftft & 5847 & 0.004 & 0147511601 & fmfmfm & 121142 & 0.038 & 0147511801 & fmfmfm & 88811 & 0.000 \\
0112372001 & ftftft & 25950 & 0.000 & 0037982401 & etftft & 11605 & 0.000 & 0037982501 & etftft & 10415 & 0.000 \\
0037982301 & etftft & 9999 & 0.008 & 0037982201 & etftft & 12699 & 0.001 & 0140160101 & fkfkfk & 41606 & 0.068 \\
0145450101 & ftftft & 8152 & 0.147 & 0145450601 & ftftft & 7978 & 0.038 & 0149890301 & ftftft  & 39996 & 0.318 \\
\hline
\end{tabular}
\label{allobstable}
}}
\end{table*}

\begin{table*}
\caption[]{Summary of the cleaned and filtered observations used in
the production of the EPIC background files, separated into the
different combinations of instrument, instrument mode and filter
used. LIVETIME is the livetime keyword value for the central CCD i.e.
corrected for periods of high-background and dead-time. For the PN
full-frame mode with the medium or thin filter, the files have been
split between the two hemispheres based on the galactic centre and
galactic anti-centre individual event pointings.}  \centering
\begin{tabular}{llllrr}
\hline
\hline
Instrument & Mode                & Filter & Location   & Number of    & LIVETIME  \\
           &                     &        & if applicable & Observations &    (s) \\ 
\hline
 MOS1      & Full-Frame          & Thin   &           &113           &    2194330\\
 MOS1      & Full-Frame          & Medium &           & 65           &     1583390\\
 MOS1      & Full-Frame          & Thick  &           & 12           &      305700\\
\hline
 MOS2      & Full-Frame          & Thin   &           &112           &     2087800\\
 MOS2      & Full-Frame          & Medium &           & 67           &     1679900\\
 MOS2      & Full-Frame          & Thick  &           & 13           &      343773\\
\hline
 PN        & Full-Frame          & Thin   & total      & 42           &      993370\\
 PN        & Full-Frame          & Thin   & anti-centre & 21          &      536349\\
 PN        & Full-Frame          & Thin   & centre     & 21          &      457021\\
 PN        & Full-Frame          & Medium & total      & 33             &     1078113\\
 PN        & Full-Frame          & Medium & anti-centre & 12           &      574272\\
 PN        & Full-Frame          & Medium & centre     & 21           &      503841\\
 PN        & Full-Frame          & Thick  &            & 10            &      237458\\
\hline
 PN        & Full-Frame-Extended & Thin   &           &62            &      825204\\
 PN        & Full-Frame-Extended & Medium &           &30            &      461575\\ 
 PN        & Full-Frame-Extended & Thick  &           & 3            &       58664\\ 
\hline
\end{tabular}
\label{finalfilestable}
\end{table*}

\section{Properties of the blank sky event files}\label{propfinal}

Some basic analysis was performed on the blank sky event files.  The
following features of these data sets should be taken into account
when using the files.

\subsection{Variations in spectral shape with count rate}

Spectra were produced selecting the time intervals for low, medium and
high count rates, using good time intervals between 10\,keV and
12\,keV. Spectra were taken within a circle of radius 8.3\arcm \,between
0.3\,keV and 8\,keV. Using MOS1, thin-filter, full-frame mode as an
example, low count rate intervals were defined as those with a count
rate less than 48 ct s$^{-1}$, medium intervals were those with a count rate
of between 48 ct s$^{-1}$ and less than 75 ct s$^{-1}$ and those of a high count
rates were those with count rates above 75 ct s$^{-1}$. The intervals were
chosen so that the total number of counts in each spectrum was similar
to aid comparison. The spectra were plotted using Xspec, using a
canned MOS1 response matrix.

\begin{figure}
\vspace{1cm}
\includegraphics[height=8cm, width=8cm, angle=270]{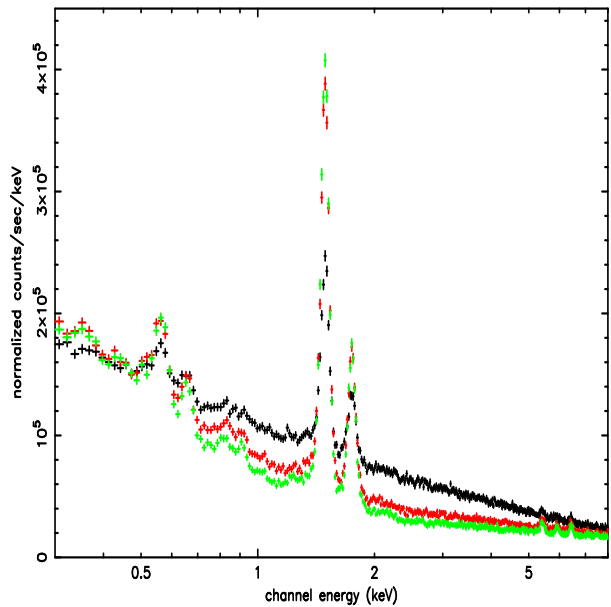}
\caption{Spectra plotted based on differences in count rate between
  0.3\,keV and 8\,keV, for MOS1, in full-frame mode using the thin filter. 
Black shows the highest count rate, red is classed
  as the medium count rate spectrum, and green the lowest count rate
  spectrum.}
\label{figcount}
\end{figure}

Figure \ref{figcount} shows that the medium and low count rate spectra
are comparable, whereas the high count rate spectrum shows enhancement
between 0.6 and 8\,keV. This effect has been seen before in studies of
background removal for cluster analysis (e.g. Temple et al. 2005).
Therefore, the user may wish to filter the BG files to an appropriate
count rate threshold, based on their own data.

\subsection{Variation of out-FOV count rate}

For each of the unfilled event lists, the out-of-field-of-view
(out-FOV) count rate, essentially the cosmic-ray induced
instrumental background, was calculated for both events with pattern
0, and events with patterns $\leq$ 12 for MOS and events with patterns
$\leq$ 4 for PN, using spectra created between the energies of 2.5\,keV
and 6.0\,keV, to avoid the instrumental lines and just show the
continuum. The out-FOV count rates representing the
quiescent particle (instrumental) background, are given in Table
\ref{crtable}.\newline

The count rates for MOS1, MOS2 and PN show an expected increase when
looking at the rates from those patterns $\leq$ 12, or 4 in the case
of PN, compared to those with pattern 0. MOS1 and MOS2 are consistent
with each other, whereas PN shows a higher count rate in comparison
with the MOS cameras. The PN camera is more sensitive, and the
differences in count rates here is indicative of this
characteristic.\newline

Differences in count rates are seen between the PN full-frame mode
event files (both thin and medium filter), selected on the basis of
their sky position $-$ i.e. Galactic centre or Galactic anti-centre
$-$ the Galactic centre count rates being somewhat larger. It was
initially thought that this difference might be due to the time of
year of the individual observations, there being some evidence that
summer observations are, on average, more heavily effected by solar
flare activity than winter observations. A higher summer-to-winter
exposure time ratio for the Galactic centre observations, compared to
the Galactic anti-centre observations, could be indicative of
relatively worse solar flaring in the Galactic centre
observations. These ratios however are not particularly different for
the two position-dependent subsets, and thus there is little evidence
that this hypothesis is correct. We note though that the numbers of
observations in each of the subsets is rather low, and small number
statistics and the actual pointings of the individual observations may
well be the cause of the observed differences. Indeed, on ignoring
outlying observations with high count rates, the ratio between centre
and anti-centre is greatly reduced.

\begin{table*}
\vspace{15mm}
\caption[]{Out-FOV and in-FOV count rates for the refilled
background event files between 2.5\,keV and 6.0\,keV.  The pattern code 
gives the patterns used, and can be
either equal to 0, $\leq$ 12 (MOS) or $\leq$ 4 (PN). For the
mode/filter combination a code is used, and for some PN files the
location dependent of the file is indicated where appropriate [f -
full-frame mode, e - full-frame-extended mode , t - thin filter, m -
medium filter, k - thick, AC - galactic anti-centre hemisphere, GC -
galactic hemisphere].}  \centering
\begin{tabular}{lccccc}
\hline
\hline
Instr & Mode/Filt./location & Pattern & Out-FOV count rate (2.5$-$6\,keV)                 & In-FOV count rate (2.5$-$6\,keV)  \\ 
      & if applicable       & code    & (ct/s/arcmin$^2$)$\times$10$^{-4}$ & (ct/s/arcmin$^2$)$\times$10$^{-4}$          \\
\hline
M1    & ft                  & 0       & 2.69 & 5.74\\
M1    & ft                  & 12      & 4.01 & 8.17\\
M1    & fm                  & 0       & 2.80 & 6.92\\
M1    & fm                  & 12      & 4.15 & 9.80\\
M1    & fk                  & 0       & 2.68 & 3.52\\
M1    & fk                  & 12      & 4.02 & 5.11\\
\hline
M2    & ft                  & 0       & 2.87 & 5.92\\
M2    & ft                  & 12      & 4.21 & 8.92\\
M2    & fm                  & 0       & 2.93 & 6.70\\
M2    & fm                  & 12      & 4.29 & 9.34\\
M2    & fk                  & 0       & 2.88 & 3.48\\
M2    & fk                  & 12      & 4.22 & 4.96\\
\hline
PN    & et                  & 0       & 20.4 & 37.2\\
PN    & et                  & 4       & 31.2 & 58.4\\
PN    & em                  & 0       & 31.3 & 62.2\\
PN    & em                  & 4       & 48.1 & 98.0\\
PN    & ek                  & 0       & 11.8 & 17.3\\
PN    & ek                  & 4       & 17.7 & 26.8\\
\hline
PN    & ft  AC              & 0       & 25.4 & 47.7\\
PN    & ft  AC              & 4       & 39.1 & 74.9\\
PN    & ft  GC              & 0       & 34.2 & 64.1\\
PN    & ft  GC              & 4       & 52.6 & 101.0\\
PN    & fm  AC              & 0       & 27.7 & 45.3\\
PN    & fm  AC              & 4       & 42.6 & 71.1\\
PN    & fm  GC              & 0       & 39.5 & 80.1\\
PN    & fm  GC              & 4       & 61.0 & 126.5\\
PN    & fk                  & 0       & 19.0 & 31.3\\
PN    & fk                  & 4       & 28.9 & 49.0\\
\hline
\end{tabular}
\label{crtable}
\end{table*}

\subsection{Variation of in-FOV count rates}

Similar to the analysis of out-FOV, in-FOV count rates were also
taken from spectra between 2.5\,keV and 6.0\,keV to avoid the
instrumental lines, using a circle of 10\arcm, with a pattern
selection as before as used for the out-FOV count rates. These
values were calculated for the refilled event files to avoid the use
of the complicated exposure maps in calculation of the areas used and
are also shown in Table \ref{crtable}. For each instrument-filter-mode
combination, the in-FOV count rates are higher than those of the
out-FOV, as expected as these in-FOV values result from both photon
and particle contributions. The MOS cameras are consistent with each
other. The PN in-FOV count rates are higher than those of the MOS.
The PN in-FOV count rates for those files split into the galactic
centre and galactic anti-centre events show the same feature as the
out-FOV, namely that the galactic centre has higher count rates
than the anti-centre. The factor by which the Galactic Centre count
rates are greater than the Galactic anti-centre count rates is larger
in-FOV than out-FOV (as expected, given that photons, soft protons
and particles contribute in-FOV, whereas only [mostly] particles
contribute out-FOV). As discussed with respect to the out-FOV
count rates however, there is no evidence that it is a time-of-year
effect that is causing these differences, but it can be explained via
statistical fluctuations and the relatively low number of observations
used.

\begin{figure}
\includegraphics[height=8cm, width=8cm, angle=270]{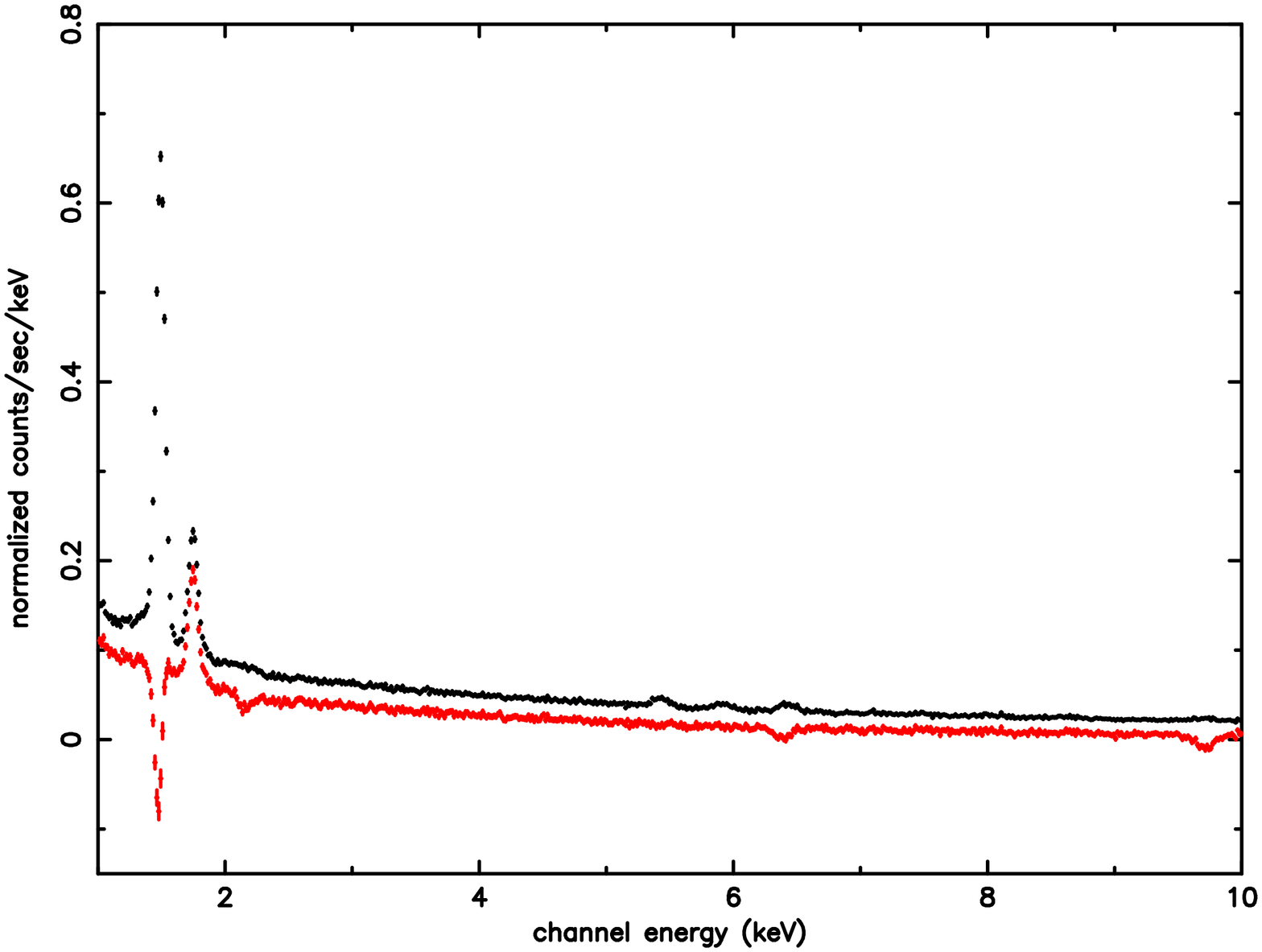}
\vspace{0.25cm}
\includegraphics[height=8cm, width=8cm, angle=270]{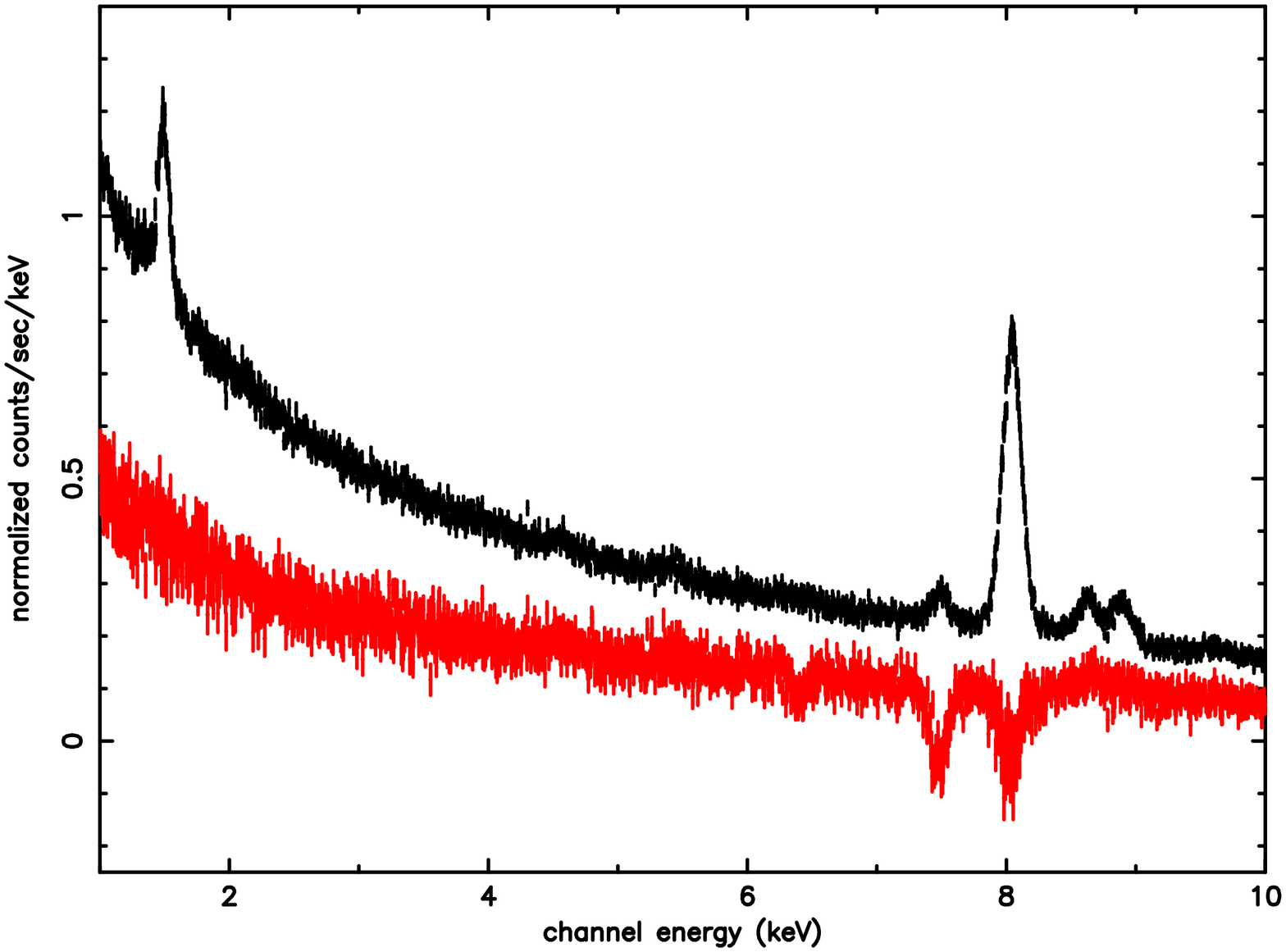}
\caption{MOS1 (top) and PN (bottom) (1.0$-$10\,keV) spectra extracted
from the full-frame, thin filter BG event files. The upper spectrum in
each panel is extracted from the in-FOV region, while the lower
spectrum shows the difference between the in-FOV spectrum and an
extracted out-FOV spectrum, scaled to account for the difference in
areas. Line features in the 'difference' spectra indicate the
differences in distribution of fluorescent line emission across the
detectors, as explained in the text.}
\label{figfovdiff}
\end{figure}

An indication of the spectral differences in- and out-FOV, and how
these differences can lead to severe problems when trying to subtract
the instrumental background, is shown in Fig. \ref{figfovdiff}. Here,
MOS1 and pn spectra are shown extracted from the in-FOV region,
together with spectra showing the difference between the in-FOV
spectrum and an extracted, area-scaled out-FOV spectrum. Strong
instrumental BG lines at Al (1.5\,keV) and Si (1.7\,keV) are seen in
the MOS spectrum, together with weaker features at Cr (5.4\,keV), Mn
(5.8\,keV), Fe (6.4\,keV) and Au (9.6\,keV). The PN spectrum shows a
strong instrumental line at Al (1.5\,keV) and a strong high-energy
line at Cu (8.0\,keV), plus weaker high-energy lines at Ni (7.4\,keV),
Cu (8.6\,keV) and Zn (9.0\,keV).

The 'difference' spectra should, ideally, show what remains once the
instrumental BG has been removed, i.e. a smooth photon BG
distribution. However this is not the case, and sharp features are
seen in the 'difference' spectra at the positions of the instrumental
lines. This is due to the distribution of the emission from the
various fluorescent lines varying across the detectors, as discussed
in Sect.\ref{bginstr}. For MOS, there is relative deficit of Al
in-FOV compared to out-FOV, hence the 'difference' spectrum shows a
trough. For Si, the situation is reversed and a peak in the difference
spectrum is observed. Similarly Au (2.2\,keV) and Fe are relatively
enhanced out-FOV. For PN, the Al line is relatively evenly distributed
across the detector, and consequently the difference spectrum shows no
peak or trough. Ni and Cu however, are very much depleted at the
centre of the detector, as discussed in Sect.\ref{bginstr}, and
strong troughs are observed in the difference spectrum.

\subsection{Variation in BG with location on sky}

\begin{figure}
\includegraphics[height=8cm, width=8cm, angle=270]{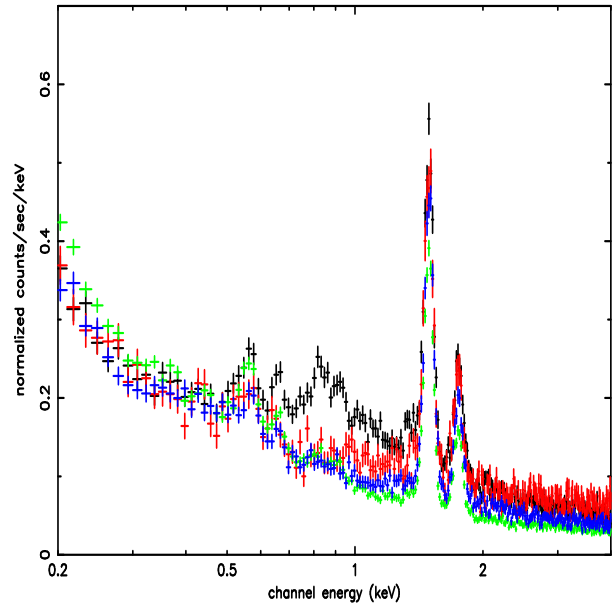}
\caption{MOS1 background spectra based on events taken from different
sky locations; centred on the galactic centre(black) and
anti-centre(red), north(green) and south(blue) galactic poles.}
\label{figloc}
\end{figure}

Background events from regions of the sky have been selected from four
coordinate centres using the script SelectRADec (see section
\ref{scripts}): the galactic centre, galactic anti-centre, the south
galactic pole and the north galactic pole. Spectra were then produced
for each region and plotted in Xspec using a canned response matrix,
for example for MOS1 thin filter, full-frame mode as shown in Figure
\ref{figloc}.

Note that the galactic centre shows a higher count rate at between
approximately 0.3\,keV to 3\,keV due to increased soft X-ray emission in
the galactic plane and centre.

\section{Using the Background files}\label{scripts}

Several useful scripts that can be used with the background event
files described in this paper, are found at the BGWG web page. The
scripts are described below.

\subsection{Script: Skycast}
 
This script is used to cast an XMM-Newton EPIC background dataset (or
indeed any EPIC event dataset) onto the sky, at the position given by
an input template event dataset (e.g. the event file for which one is
interested in producing a background). This is useful as the user is
more likely to be working in sky coordinates than instrument
coordinates.
	 
\subsection{Script: SelectRADec}
This script can be used to select events from a certain area of the
sky, specifying a right ascension and declination (J2000), and the
maximum distance from this position one is prepared to consider. A
final event file and exposure map is produced. This may be useful to a
user whose is concerned about the dependence of the background with
location, as shown in Figure \ref{figloc}.

\subsection{Script: BGrebin} 
This script is used to rebin and re-project the provided exposure maps
onto the sky to the spatial scale and sky position of a user-input
image.  This is useful when wanting to work in sky coordinates and to
a different spatial scale than the 4\arcs scale used in the
production of the exposure maps.

\subsection{Using the Background files: Reliability}\label{testing}
In this final subsection on the use of the BG files, we have attempted
to provide a gauge of the reliability of the Blank Sky files. To
achieve this, and to mirror how these files might be used by the
community, we have performed a typical BG analysis that a general user
might perform, but on separate {\em subsets} of the same blank sky
event files, and compared the results obtained, i.e.\,employing a
'cross-validation'-like technique. For each analysis, two large
time-separated sub-event files were extracted from a particular BG
event file, each covering 40$-$50\% of the original exposure
time. These two event files were then screened (further) for times of
medium-high soft proton flares, as a user would wish to do (the event
files having been earlier screened for times of high soft proton
flares $-$ see Sect.3). For each flare-filtered sub-event file, a
'source' spectrum was extracted from a 6\arcm radius circular in-FOV
central region of the instrument. A 'background' spectrum was also
extracted from the out-FOV 'corners' of the detector. Here then, the
'source' spectrum should contain (predominantly) both photon and
particle components of the EPIC BG, whereas the 'background' spectrum
should contain just the particle component of the EPIC
background. Correct analysis and subtraction of one from the other
should yield just the extragalactic photon X-ray background.

For the MOS cameras, the out-FOV regions of the detectors are
quite large, and hence reasonable instrumental BG statistics can be
obtained from these areas (in actuality, we used the MOS out-FOV
regions as defined in de Luca \& Molendi (2004)). For the pn however,
the situation is much worse, as the out-FOV regions are very much
smaller and close to the noise of the pn readout. The very limited
statistics obtained from the pn out-FOV regions make instrumental BG
subtraction very problematic. Furthermore, for the pn, the response
varies along each CCD due to charge transfer inefficiency effects,
making any determination of a linear response with position and radial
vignetting a very complicated problem (e.g.\,Lumb et al 2002). The
recent addition of long-exposure stacks of closed filter full-FOV pn
data (dominated by the instrumental BG) to the official ESA web pages
(\textit{http://xmm.vilspa.esa.es/external/}\newline
\indent \indent \textit{xmm\_sw\_cal/background/index.shtml.} will
help in the future to allow a much better determination and
subtraction of the in-FOV instrumental BG component. This is beyond
the scope of the present paper however, and we have here concentrated
on the MOS data.

The analysis using the large, MOS, thin filter, full-frame BG event
files is described here. Once the spectra were formed, as above,
appropriate RMF response matrices and ARF auxiliary files were created
using XMM-Newton SAS 7.0 and the source spectral channels were binned
together to give a minimum of 20 counts per bin. Spectral fitting was
performed using Xspec, incorporating a photo-electric absorption
times power-law (\em{wabs}$\times$\em{power-law}) model. The hydrogen
column density of the absorption component was fixed in each case to
the exposure-weighted average of the hydrogen column densities of the
component observations (see Sect.3.2). The power-law spectral index
and normalization were allowed to go free. Fitting was performed in
the energy range 2.5$-$6.0\,keV, i.e.\, concentrating on the line-free
continuum area of the spectrum. The spectral analysis results are
summarized in Table \ref{relitable}. Tabulated, both for MOS1 and MOS2, are the
best-fitting power-law spectral index and normalization for each
'half' of the original full-size thin filter, full-frame BG event
files. Errors are 1$\sigma$ for one interesting parameter. As can be
seen, the fitted indices and normalizations for the two halves are
consistent with one another, both for MOS1 and for MOS2. Furthermore,
the values obtained for MOS1 are consistent with the values obtained
for MOS2. And finally, the actual values of power-law index obtained
are consistent with the generally accepted value for the power-law
index of the true extragalactic X-ray background ($\approx1.42$;
e.g.\,Lumb et al 2002). This analysis therefore, indicates that the BG
event files presented here are consistent (both with each other and
with other external measures of the X-ray BG) and reliable (both
between files and internally within individual files).

\begin{table}
\vspace{15mm}
\caption[]{Results of spectral fitting to the two subsections of the
MOS thin filter, full-frame mode background files, fitted using a
wabs$\times$power-law model (see text).} \centering
\begin{tabular}{lccccc}
\hline
\hline
Instr & Subsection & Photon Index  & Normalisation       \\ 
      &            &               & ($\times$10$^{-5}$) \\
\hline
M1    & 1          & 1.37$\pm$0.15 & 7.64$\pm$1.50\\
M1    & 2          & 1.45$\pm$0.15 & 10.5$\pm$1.90\\
M2    & 1          & 1.42$\pm$0.07 & 12.2$\pm$1.17\\
M2    & 2          & 1.50$\pm$0.19 & 9.62$\pm$2.30\\
\hline
\hline
\end{tabular}
\label{relitable}
\end{table}

\section{Conclusions and future developments}\label{concl}

We have described in detail the components that make up the XMM-Newton
EPIC background and why a good understanding of this background is
important. We have explained the steps behind the production of blank
sky event lists that are available for the general user via the
XMM-Newton EPIC background working group website, and are useful for
background analysis. We have presented several interesting features of
these files including variations in spectral shape with count rate and
variations in count rate between out-FOV and in-FOV areas. Finally we
have presented software that can be used with the blank sky event
files and associated exposure maps, also available via the background
working group web pages.

It is hoped that in the near future more 2XMM datasets will become
available that fulfil the criteria previously described for the
background sets. These will then be filtered as appropriate and merged
with the existing datasets. We also hope to improve the
ghosting procedure as described in Section \ref{ghosting} to
incorporate the inclusion of bad pixels and columns, and to provide a
means to select background events based on their component observation
exposure time, as described in Section \ref{propobs}. A tool to
select sections of the event files as based on count rates may become
available in the near future, to accompany the tools already
available.

Further releases of modified datasets will be announced via the URL:
\newline \textit{htpp://xmm.esac.esa.int/external/}\newline \indent
\textit{xmm\_sw\_cal/background/index.shtml}.

\section{Acknowledgements}

We would like to acknowledge the work of, and thank, all other members
and colleagues of the EPIC background working group: M. Ehle,
M. Freyberg, M. Kirsch, K. Kuntz, A. Leccardi, S. Molendi, W. Pietsch
and S. Snowden. We thank and acknowledge the anonymous referee
for helpful and useful comments that have improved this paper.

\end{document}